\documentclass[3p,times,sort]{elsarticle}

\usepackage[T1]{fontenc}
\usepackage{url}
\usepackage{xurl}
\usepackage{pifont}
\usepackage{amssymb,amsmath}
\usepackage{graphicx}
\usepackage{lineno}
\usepackage{siunitx}
\usepackage{placeins}
\usepackage{booktabs}
\usepackage{subcaption}

\begin{document}

\begin{frontmatter}

  \title{The Sim-to-Real Gap in MRS Quantification: A Systematic Deep Learning Validation for GABA}

  \author[cf]{Zien Ma}
  \author[sw]{S.\ M.\ Shermer}
  \author[cf]{Oktay Karaku\c{s}}
  \author[cf]{Frank C.\ Langbein\corref{cor1}}
  \ead{LangbeinFC@cardiff.ac.uk}
  \cortext[cor1]{Corresponding author.}

  \affiliation[cf]{organization={School of Computer Science and Informatics, Cardiff University},
    city={Cardiff},
    postcode={CF24 4AG},
  country={United Kingdom}}
  \affiliation[sw]{organization={Faculty of Science and Engineering, Swansea University},
    city={Swansea},
    postcode={SA2 8PP},
  country={United Kingdom}}

  \begin{abstract}
    Magnetic resonance spectroscopy (MRS) is used to quantify metabolites \emph{in vivo} and estimate biomarkers for conditions ranging from neurological disorders to cancers. Quantifying low-concentration metabolites such as GABA ($\gamma$-aminobutyric acid) is challenging due to low signal-to-noise ratio (SNR) and spectral overlap. We investigate and validate deep learning for quantifying complex, low-SNR, overlapping signals from MEGA-PRESS spectra, devise a convolutional neural network (CNN) and a Y-shaped autoencoder (YAE), and select the best models via Bayesian optimisation on $10{,}000$ simulated spectra from slice-profile-aware MEGA-PRESS simulations. The selected models are trained on $100{,}000$ simulated spectra. We validate their performance on $144$ spectra from $112$ experimental phantoms containing five metabolites of interest (GABA, Glu, Gln, NAA, Cr) with known ground truth concentrations across solution and gel series acquired at $\SI{3}{T}$ under varied bandwidths and implementations. These models are further assessed against the widely used LCModel quantification tool. On simulations, both models achieve near-perfect agreement (small MAEs; regression slopes $\approx 1.00$, $R^2 \approx 1.00$). On experimental phantom data, errors initially increased substantially. However, modelling variable linewidths in the training data significantly reduced this gap. The best augmented deep learning models achieved a mean MAE for GABA over all phantom spectra of $0.151$ (YAE) and $0.160$ (FCNN) in max-normalised relative concentrations, outperforming the conventional baseline LCModel ($0.220$). A sim-to-real gap remains, but physics-informed data augmentation substantially reduced it. Phantom ground truth is needed to judge whether a method will perform reliably on real data.
  \end{abstract}


\end{frontmatter}

\section{Introduction}\label{sec:introduction}

Magnetic resonance spectroscopy (MRS) is a non-invasive technique for quantifying metabolite concentrations \emph{in vivo}, providing insights into cellular metabolism that can support diagnosis and monitoring across neurological and oncological diseases~\cite{Blaszczyk2016, Madeira2018, Rideaux2021, Yuan2016, ZHANG2022}. A biomarker of particular clinical interest is $\gamma$-aminobutyric acid (GABA), the primary inhibitory neurotransmitter in the brain, whose dysregulation is implicated in numerous psychiatric and neurological disorders~\cite{Nuss2015, Luscher2011, Puts2012, Bollmann2015, Edden2012, Jewett2022, Blaszczyk2016}. Accurate quantification of low-concentration metabolites such as GABA is, however, challenging: the weak GABA signal is partially obscured by stronger resonances from more abundant compounds such as N-acetylaspartate (NAA) and creatine (Cr), leading to substantial spectral overlap. Edited MRS techniques such as MEGA-PRESS (Mescher-Garwood point-resolved spectroscopy editing) are therefore commonly employed to isolate the GABA signal~\cite{MEGAPRESS1998, MEGAPRESS1996, Mullins2014}. While spectral editing improves specificity, the required subtraction of edit-OFF and edit-ON (hereafter OFF and ON) acquisitions reduces the signal-to-noise ratio (SNR) and may introduce artefacts, creating additional challenges for reliable, unbiased quantification.

Deep learning (DL) has been applied to address these problems, potentially improving accuracy and reducing expert-driven parameter tuning~\cite{Dias2024}. In practice, however, DL methods face a major obstacle: the ``sim-to-real'' gap between simulated training data and experimental measurements. Training robust models requires large, labelled datasets\@. \emph{In vivo} data are costly to acquire at scale, subject to ethical and logistical constraints, and crucially lack ground-truth metabolite concentrations, precluding fully supervised training and rigorous evaluation~\cite{GABA2023, Consensus2019, Rideaux2021}. Phantom datasets provide known concentrations but are expensive and time-consuming to prepare, calibrate (e.g.\ pH, temperature, relaxation), and scan under multiple conditions; covering all concentration combinations, linewidths, and sequence variants would be impractical~\cite{BENCHMARK,Saleh2019}. As a result, most DL models are trained and selected on large collections of simulated spectra~\cite{FID-A2017, MRSNET, Rizzo2022}.

Simulations typically assume idealised hardware, controlled acquisition parameters, and simplified baselines. Experimental spectra, instead, show variations and artefacts from scanner-specific implementations, B0/B1 inhomogeneities, imperfect water suppression, and subtraction-related baseline distortions~\cite{Mikkelsen2017, Mikkelsen2019, Saleh2019}. Models optimised exclusively on simulated data can therefore overfit to unrealistic training distributions and exhibit substantial estimation bias when applied to experimental spectra. Recent studies have highlighted this concern, reporting strong performance on simulations but degraded accuracy and calibration under domain shift~\cite{RizzoMICCAI2022, Rizzo2022}.

In this work, we directly address this validation challenge through a systematic investigation of DL-based quantification of GABA and related metabolites from MEGA-PRESS spectra. While GABA is our primary target, we simultaneously quantify NAA, Cr, glutamate (Glu), and glutamine (Gln), which are integral to brain metabolism and function~\cite{Metabolite2013}. Building on MRSNet~\cite{MRSNET}, we pursue three main objectives. Firstly, we develop two complementary architectures for multi-metabolite regression: a convolutional neural network (CNN) that captures local spectral features and a Y-shaped autoencoder (YAE) that learns a denoised latent representation. Secondly, we perform systematic model selection via Bayesian optimisation on a slice-profile-aware simulated dataset (five-fold cross-validation on $10{,}000$ spectra) to identify the best performing configurations of these models, and include several established architectures from the literature as comparative baselines (using their published configurations with minimal adaptation to our MEGA-PRESS pipeline). Thirdly, we assess the performance of the best DL models by validating them on $144$ spectra from $112$ experimental phantoms with known ground-truth concentrations (solutions and tissue-mimicking gels acquired at $\SI{3}{T}$ containing GABA, Glu, Gln, Cr, NAA and no macromolecule background or lipid signal), and comparing their performance to the widely used LCModel tool (applied to OFF spectra as it provided the better quantification results on the phantom data). For the baseline (fixed-linewidth, non-augmented) models, across all $144$ phantom spectra, mean MAE over all spectra for GABA was $0.161$ (MRSNet-YAE), $0.203$ (MRSNet-CNN), $0.167$ (FCNN), and $0.220$ (LCModel-OFF).

Our results show that both architectures achieve near-perfect agreement with ground truth on simulated data. On experimental phantoms, initial models showed a substantial sim-to-real gap. However, by incorporating realistic variability in spectral linewidths into the training data, we significantly improved robustness. The augmented models outperformed the conventional LCModel baseline for GABA and Glu; with linewidth augmentation, the best DL models achieved GABA MAE over all phantom spectra $0.151$ (YAE) and $0.160$ (FCNN), outperforming LCModel-OFF ($0.220$), even if the sim-to-real gap could not be closed. Physics-informed simulation such as linewidth augmentation is, hence, important for DL methods to generalise from simulation to experiment. We argue that validation on phantoms with known concentrations should precede \emph{in vivo} use of any new quantification method.

Section~\ref{sec:relatedwork} reviews conventional and machine learning-based quantification methods in MRS\@. Section~\ref{sec:data} details the simulated and experimental datasets and preprocessing. Section~\ref{sec:archs} describes our DL architectures, and Section~\ref{sec:selection} outlines the Bayesian optimisation-based model selection. Section~\ref{sec:experiments} presents the experimental results; Section~\ref{sec:conclusion} concludes.

\section{Related Work}\label{sec:relatedwork}

The quantification of metabolites from MRS spectra, particularly low-concentration compounds like GABA, has motivated a wide range of analytical methods~\cite{GABA2021, GABA2023, GABA2018}. These fall into two main paradigms: (i) conventional model-based fitting with explicit basis functions and (ii) data-driven machine learning, including deep learning (DL). We focus on methods for edited spectra such as MEGA-PRESS and summarise their main developments and limitations, in particular interpretability, uncertainty quantification, and behaviour under domain shift.

\subsection{Conventional Quantification Methods}

Conventional quantification in MRS is dominated by peak fitting and spectral basis set methods. Peak fitting models spectral peaks using analytical lineshapes (e.g., Gaussian, Lorentzian, Voigt), adjusting parameters like amplitude and phase to match the data and estimating metabolite concentrations from peak areas. While some tools, such as LWFIT~\cite{BENCHMARK}, use model-free integration over fixed frequency intervals for robustness, others, such as GANNET~\cite{GANNET} and its successor Osprey~\cite{Osprey2021}, specialise in edited spectra like MEGA-PRESS by fitting Gaussian models to target peaks. Although effective for well-separated signals, peak fitting accuracy is often compromised by spectral overlap and baseline distortions.

Basis set fitting methods model the acquired spectrum as a linear combination of basis spectra, which are pre-acquired from phantom experiments or generated from quantum-mechanical simulations. Using fitting algorithms such as constrained nonlinear least-squares, these methods determine the relative contributions of each metabolite. A variety of such tools exist, operating either in the frequency (e.g., LCModel~\cite{LCModel1993, LCModel2001}, INSPECTOR~\cite{INSPECTOR}, VESPA~\cite{VESPA}, JMRUI~\cite{JMRUI}, AQSES~\cite{AQSES}, AMARES~\cite{AMARES}) or time domain (e.g., QUEST~\cite{QUEST}, TARQUIN~\cite{TARQUIN2006, TARQUIN2011}), as reviewed in~\cite{MRSreview}. Recent toolboxes also offer Bayesian posterior estimates over concentrations (e.g., FSL-MRS~\cite{FSL-MRS}), improving uncertainty characterisation relative to point estimates. Despite their widespread use, these methods face several challenges. Simulated basis sets may not fully capture the nuances of experimental spectra~\cite{Kreis2012}, and performance can degrade with low signal-to-noise ratio or increased spectral linewidth~\cite{Zoellner2020, Zoellner2021, Craven2024}. Experimentally acquired basis sets can improve accuracy, but their generation is laborious and requires significant human expertise for preprocessing and parameter tuning, introducing operator-dependent variability~\cite{Consensus2019}. Furthermore, challenges such as macromolecule contamination and ensuring consistency across different modelling choices persist~\cite{Craven2022, DaviesJenkins2024}.

\subsection{Machine Learning Methods}

Machine learning and particularly Deep learning (DL) have been proposed as an alternative that can automate analysis and reduce dependence on manual tuning and operator-dependent variability. Models trained on large-scale simulated datasets have been shown to learn relevant features and predict metabolite concentrations directly from spectral data~\cite{RizzoMICCAI2022, Rizzo2022, MRSNET, QNet2024, CloudBrain-MRS2024, Das2017, Dziadosz2023, Hatami2018, Lee2019, Lee2022}; other work uses DL for denoising before conventional linear least-squares (LLS) or linear combination model (LCM) quantification~\cite{QNet2024, CloudBrain-MRS2024, Lee2019, Lee2022, SHAMAEI2023}. We group existing DL-based quantification strategies into three categories. A review of machine learning applications in MRS can be found in~\cite{Vandesande2023}.

\textbf{Direct Regression Approaches:} The most direct application of DL frames quantification as an end-to-end regression task, where a network, typically a CNN, learns to map raw spectral data directly to metabolite concentrations~\cite{RizzoMICCAI2022, Rizzo2022, Hatami2018}. Variations on this theme include using handcrafted wavelet scattering features to provide more robust inputs to a shallow regression network~\cite{Shamaei2021}. These models are computationally efficient and can be designed to provide uncertainty estimates. However, because they do not explicitly model the physics of spectral formation, they can lack interpretability and are prone to overfitting, especially in low-SNR regimes. Their estimation bias is often unknown, as most have not been validated against experimental data with ground-truth concentrations.

\textbf{Physics-Informed and Hybrid Models:} To improve interpretability and robustness, other approaches integrate physical knowledge of spectral composition into the DL framework. One strategy involves using a CNN to predict a clean, metabolite-only spectrum, which is then quantified using a fixed, non-trainable solver such as multivariable linear regression against a predefined basis set~\cite{Lee2019, Lee2022, Q-MRS2024, HuangTW2024}. A key limitation here is that errors from the quantification step cannot be backpropagated to train the feature extractor. More advanced methods overcome this by incorporating a fully differentiable solver into the network architecture. For example, Q-Net~\cite{QNet2024} embeds a differentiable least-squares solver, while PHIVE~\cite{PHIVE2025} integrates a differentiable LCM into a variational autoencoder, enabling end-to-end training and robust uncertainty estimation. These models more tightly link representation learning with the quantification task, but are more complex to train and require access to accurate basis sets.

\textbf{Heuristic and Partially Differentiable Models:} A third category of methods uses DL for feature extraction or parameter estimation but relies on fixed or heuristic components for the final quantification. For instance, some models use a denoising autoencoder to learn a robust spectral representation but decode it using a fixed LCM with a standard basis set~\cite{SHAMAEI2023}. Other encoder-decoder designs extend this approach. For example, Zhang \emph{et al.} proposed a model using WaveNet blocks and an attention-based GRU to learn a robust latent representation from multi-echo JPRESS data, from which it simultaneously predicts concentrations, reconstructed FIDs, and phase parameters in an end-to-end manner~\cite{encdec2023}. Others use a CNN to estimate spectral parameters (e.g., peak locations and widths) and then reconstruct the spectrum using a fixed but differentiable physical model, such as a sum of Lorentzian functions~\cite{Gurbani2019}. A different approach is taken by NMRQNet~\cite{NMRQNet2023}, which uses a recurrent network to predict spectral parameters and then refines them with a non-differentiable stochastic optimisation algorithm. While these methods can enforce physical plausibility, the separation between the learned and fixed components prevents global error minimization and limits end-to-end optimisation.

Conventional LCM methods remain interpretable and can provide uncertainty estimates but depend on accurate basis sets and can be sensitive to linewidth and baseline~\cite{Zoellner2021, Craven2024}. Direct-regression DL is fast and flexible but often lacks calibrated uncertainty and has rarely been validated on experimental ground truth~\cite{Rizzo2022, Dziadosz2023}. Physics-informed and differentiable hybrids improve interpretability and gradient flow~\cite{QNet2024, PHIVE2025} but still rely on basis accuracy and remain under-validated on phantoms. Recent reviews of ML for proton MRS (e.g.\ Vandesande \emph{et al.}~\cite{Vandesande2023}) describe progress in denoising, quantification, and uncertainty, and note remaining concerns about generalisation and bias under domain shift~\cite{Dziadosz2023}. Many DL studies report strong performance on simulations or on \emph{in vivo} data without ground truth; few provide validation on experimental phantoms with known concentrations.

\subsection{Limitations of Existing Work and Our Contribution}

Conventional and machine learning methods suffer from a lack of validation against known ground truth. Many DL-based studies rely exclusively on simulated data for evaluation, using pre-set concentrations as ground truth~\cite{RizzoMICCAI2022, Rizzo2022, Hatami2018, Shamaei2021}. Others use \emph{in vivo} data, which lacks ground truth, and instead compare results to established methods such as LCModel~\cite{QNet2024, Das2017, Lee2019, Lee2022}. As recent work has highlighted, models validated solely on synthetic data often overfit and exhibit significant estimation bias when applied to real-world spectra~\cite{RizzoMICCAI2022, Rizzo2022}. Thus, neither approach provides a true measure of a model's practical utility.

Our work addresses this validation gap. We evaluate DL models for MEGA-PRESS quantification systematically: Bayesian model selection on simulations, then validation on $144$ phantom spectra with known concentrations. We develop two distinct architectures, a CNN and a novel Y-shaped autoencoder (YAE), and optimise them through Bayesian hyperparameter search on a large simulated dataset. The main focus is validation of the selected models on experimental phantom spectra and comparison with LCModel (Section~\ref{sec:data}). We compare our models to ground truth and to the widely used LCModel to assess the capabilities and limitations of deep learning for MEGA-PRESS quantification. Without validation on phantoms with known concentrations, one cannot judge whether a new method will perform reliably on real data.

\section{Simulation, Phantom Experiments, and Evaluation Metrics}\label{sec:data}

Our study focuses on quantifying five metabolites (GABA, Glu, Gln, NAA, and Cr) from edited spectra acquired with the MEGA-PRESS pulse sequence, which yields OFF and ON acquisitions and their difference (DIFF\@; the ON minus OFF difference spectrum). GABA is the primary target; Glu and Gln are key excitatory neurotransmitters; NAA and Cr serve as prominent reference metabolites whose distinct peaks support calibration and relative quantification. Concentrations are often reported relative to NAA or Cr. Below, we describe the simulated and experimental datasets, preprocessing, and evaluation metrics.

\subsection{Simulated Datasets}\label{sec:data_sim}

To train and validate the models, we generated simulated spectra by taking a weighted sum of simulated basis spectra for each metabolite. OFF and ON spectra for each metabolite were generated using custom MATLAB code and the FID-A toolbox~\cite{FID-A2017}. The simulations are based on Hamiltonian models for the molecules of interest and involve solving the time-dependent Schr{\"o}dinger equation for the MEGA-PRESS pulse sequence, calculating the predicted time-domain echo signals, and adding line broadening to enhance realism. For some molecules, competing Hamiltonian models exist, including the Govindaraju, Kaiser, and Near models~\cite{Gov2000, Kaiser2008, Near2013}. In this work, we used the Near model for GABA because the spectra it produces most closely match our experimental data, and there is a consensus that it is the preferred model~\cite{Consensus2019}.

\begin{figure}
  \centering
  \subfloat[Sequence Diagram for MEGA-PRESS]{\includegraphics[width=0.5\linewidth]{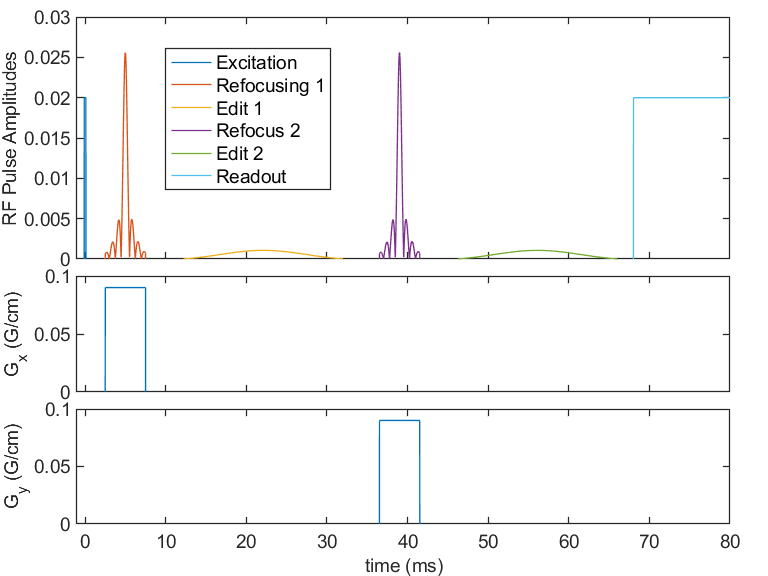}}
  \subfloat[Simulation Grid (2D) for cubic voxel with $a=\SI{3}{cm}$]{\includegraphics[width=0.45\linewidth]{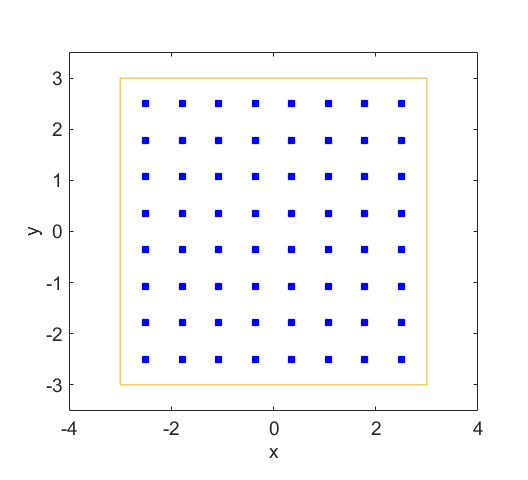}}
  \caption{The sequence diagram (a) shows the RF and gradient pulses with actual pulse shapes and timings used in the simulations. The initial excitation pulse is modelled as an ideal (instantaneous) slice-selective $90^\circ$ pulse and the corresponding slice selection gradient $G_z$ is therefore omitted. The excitation pulse excites a slice of thickness $\SI{3}{cm}$ perpendicular to the $z$-axis. The refocusing pulses refocus the magnetisation of $\SI{3}{cm}$ thick slabs perpendicular to the $x$ and $y$ axis, respectively, to define the localised voxel. What differentiates the MEGA-PRESS sequence from the standard PRESS sequence is the presence of two $\SI{20}{ms}$ frequency-selective Gaussian editing pulses (yellow and green) at $\SI{1.9}{ppm}$ for the ON acquisition. For the OFF spectra the editing pulses could in principle be omitted, but the simulation follows the experimental implementation where editing pulses at $\SI{7.5}{ppm}$, which have no effect on the metabolites of interest, are applied instead. The readout of the signal starts at $\SI{68}{ms}$ as indicated. Experimentally, it can last over $\SI{1}{s}$, depending on the dwell time and number of samples acquired. For a bandwidth of $\SI{2000}{Hz}$, the dwell time is $\SI{0.5}{ms}$ and acquiring $N=2048$ samples, a typical signal length, would therefore require $2048 \times \SI{0.5}{ms} = \SI{1.024}{s}$.  The readout block in the diagram is truncated at $\SI{80}{ms}$ for clarity, to show the RF pulses and timings. To account for imperfect slice profiles of the refocusing pulses, the spectra are simulated on a spatial grid (b) and the average over all positions is calculated.}\label{fig:MEGAPRESS}
\end{figure}

Spatial localisation for MEGA-PRESS spectra is commonly achieved using a technique similar to PRESS (Point RESolved Spectroscopy). This process begins with an initial radiofrequency pulse that excites a slab-shaped region, providing localisation in one dimension. This is followed by two subsequent refocusing pulses, each designed to selectively refocus spins within a slab perpendicular to the previously excited slab and to each other. Consequently, only spins in the intersection of these three orthogonal slabs experience the complete sequence and contribute to the final signal. Crusher gradients further enhance voxel localisation. Slice-selective excitation is achieved by applying finite-bandwidth modified sinc pulses concurrently with gradients. Many simulations neglect this, implementing excitation and refocusing pulses as ideal $90^\circ$ and $180^\circ$ rotations, respectively. However, this ignores the non-uniformity of excitation and refocusing pulse profiles, especially for large voxel sizes. To obtain more realistic spectra, we performed simulations on a 2D spatial grid to more accurately model the excitation profiles of the refocusing pulses, thereby improving the realism of the simulated spectra. Finally, phase cycling was implemented, cycling over two phases ($0^\circ$, $90^\circ$) each for the first editing and the refocusing pulses, and four phases ($0,90,180,270$) for the second editing pulse, resulting in $2 \times 2 \times 2 \times 4 = 32$ simulation runs for each point on the spatial grid, and a total of $32 \times 8 \times 8 = 512$ simulations for each metabolite spectrum. To simulate both the ON and OFF spectra for each metabolite therefore requires $1{,}024$ simulation runs. See Figure~\ref{fig:MEGAPRESS} and \texttt{mrsnet/simulators/fida/run\_custom\_simMegaPRESS\_2D.m} in~\cite{MRSNET-code} for details.

The simulation produces a complex time-domain signal corresponding to the time-evolution of the $x$ and $y$ components of the transverse magnetisation, which is Fourier-transformed to obtain a spectrum. In practice, the time-domain signal is multiplied by an exponential envelope to simulate signal attenuation due to $T_1$ and especially $T_2$ relaxation effects. The decay rate of the exponential function controls the linewidth of the simulated spectra. Lorentzian fits of the NAA peak in $144$ experimental MEGA-PRESS OFF spectra yielded a median linewidth of just under $\SI{2}{Hz}$ and a mean just over $\SI{2}{Hz}$ (full width at half maximum, FWHM). Based on this, we chose a linewidth of $\SI{2}{Hz}$ FWHM for the simulated basis spectra as a realistic but slightly conservative value. Estimated linewidths across the $144$ phantom spectra span approximately $\SI{1}{Hz}$ to $\SI{10}{Hz}$ FWHM (see the sim-to-real repository~\cite{results-sim2real}); we use this range when assessing variable linewidth augmentation in Section~\ref{sec:linewidth_aug}. Consistently, our sim-to-real analysis on the phantom series~\cite{results-sim2real} indicates that simulations with a fixed $\SI{2}{Hz}$ FWHM basis provide the closest overall match in spectral magnitude and linewidth to the phantom data, and allowing the simulated linewidth to vary did not materially reduce the sim-to-real gap in these spectral similarity metrics. In Section~\ref{sec:linewidth_aug}, we show that linewidth augmentation can improve quantification performance.

\subsection{Concentration Sampling and Noise Injection}

We adopt the same strategy for generating synthetic spectra as in our previous work~\cite{MRSNET}, where training and validation samples are constructed by taking linear combinations of individual metabolite signals from a given basis set. Each metabolite is assigned a scaling factor within the range $[0,1]$, representing its relative concentration. These concentration values are sampled using a Sobol sequence: a low-discrepancy, quasi-Monte Carlo method that ensures uniform coverage of the high-dimensional concentration space even with a relatively small number of samples. Sobol sampling gives more uniform coverage of the concentration space than random sampling, which helps the models cope with diverse spectral mixtures.

Time-domain Gaussian noise with zero mean and standard deviation $\sigma$ sampled uniformly in $[0,0.03]$ is added to all simulated spectra, whose signal amplitudes are normalised so the maximum spectral peak equals $1.0$. This noise level reflects realistic scanner-induced variability and was previously shown to closely approximate noise observed in experimental phantom spectra~\cite{MRSNET,BENCHMARK}. The range was guided by expert analysis of spectral regions that contain no identifiable metabolic signal, capturing acquisition-related baseline fluctuations. Beyond mimicking experimental conditions, randomised noise both improves realism and reduces overfitting to noise-free inputs. We verified the interval on our $144$ phantom spectra by estimating $\sigma$ from signal-free regions only in the phantom--simulation residual: median $\sigma \approx 0.014$, $\approx80\%$ within $[0,0.03]$. The $30$ spectra with $\sigma > 0.03$ come from the gel series E4, E9, and E11 (different bandwidths or reacquired later) and the solution series E6; E6 and E11 are the noisiest. These series were included to test performance across SNR\@.

For the main training runs reported in this paper, the simulated data use the above setup: slice-profile-aware basis spectra with fixed $\SI{2}{Hz}$ linewidth, Sobol-sampled concentrations, time-domain Gaussian noise in $[0, 0.03]$, and the same B0 alignment, ppm range, and amplitude normalisation as in the common export pipeline (Section~\ref{sec:preproc}). We did \emph{not} apply explicit augmentation for frequency drift, higher-order phase errors, variable linewidth beyond the fixed simulation value, synthetic rolling baselines, or non-Gaussian noise. The impact of more aggressive augmentation, specifically using varied linewidths, is explored in Section~\ref{sec:linewidth_aug} and further discussed in Section~\ref{sec:conclusion}.

\subsection{Experimental Dataset}

In addition to validation on simulated spectra, we evaluate our models' performance on experimental spectra from test objects with known metabolite concentrations, ranging from buffered metabolite solutions to tissue-mimicking gel phantoms. These phantoms contain only the specified metabolites, and no macromolecule (MM) or lipid signal, unlike \emph{in vivo} spectra.

Several sets of experiments were conducted, ranging from buffer solutions with a few metabolites to tissue-mimicking gel phantoms with several relevant metabolites. All phantoms were prepared in-house at the Institute for Life Science at Swansea University. For the solution datasets, the general procedure was to prepare a pH-neutral phosphate buffer solution and add fixed amounts of various metabolites to obtain a base solution. Some of the base solution was then used to prepare ``spiked'' solutions with high concentrations of a single metabolite, typically GABA, our primary interest. The solution series were then generated by incrementally removing a small amount of solution from the phantom and replacing it with the same volume of a spiked solution. This procedure allows for increasing the concentration of a single metabolite in small increments without changing the baseline concentration of others, providing more precise control over concentrations than if the solutions were prepared independently. After each addition of spiked solution, a new OFF and ON spectrum was acquired.

The gel phantoms were obtained similarly by preparing base solutions and adding small amounts of spiked metabolite solutions to vary concentrations. However, separate gel phantoms were prepared for each combination of concentrations, as incremental addition is not feasible for gels. To create gels, a gelling agent (Agar Agar, $\SI{1}{g}$ per $\SI{100}{ml}$) was added to each solution, and the mixture was heated to approximately $\SI{95}{\celsius}$. The hot solutions were then transferred to suitable molds and allowed to cool to room temperature and solidify before scanning.

The composition of the phantoms was intentionally varied in complexity. In total, $144$ spectra were obtained from $112$ phantoms (some phantoms were scanned multiple times---e.g.\ E4 and E9---yielding more spectra than physical objects). Solution series E1 had the simplest composition, with only fixed concentrations of NAA and Cr and increasing amounts of GABA, and no Glu or Gln. Solution series E2 was similar but deliberately miscalibrated with a low pH to test the models' ability to cope with miscalibrated data. Solution series E3 was similar to E1 but included fixed amounts of Glu and Gln. Series E4 consisted of gel phantoms with fixed amounts of NAA, Cr, Glu, and Gln, and varying amounts of GABA, with concentrations similar to E3. The gel phantoms were scanned four times. The first two rounds (E4a and E4b) were acquired on the same day with the same sequence but two different acquisition bandwidths. The phantoms were scanned again with the same sequence and bandwidths a week later to acquire more data and assess any deterioration over time (E4c, E4d). E6 and E7 were solution series similar in composition to E3, while E8, E9, E11, and E14 were gel phantoms similar to E4. Not all available experimental series were included, as some involved additional metabolites for other projects. A summary of the metabolite concentrations for the experimental series is in Table~\ref{BenchmarkExperiment}, and further details on phantom design can be found in~\cite{BENCHMARK}.

All spectra were acquired on a Siemens MAGNETOM Skyra 3T MRI system at Swansea University's Clinical Imaging Unit. Datasets E1 to E4 used the Siemens WIP MEGA-PRESS implementation for GABA editing, while the remaining datasets (E6, E7, E8, E9, E11, and E14) were acquired with a widely used implementation from the University of Minnesota~\cite{CMRR}. As GABA is our main metabolite of interest, most experimental series kept the concentrations of NAA, Cr, Gln, and Glu constant while gradually increasing GABA concentration, except for E9, where all metabolite concentrations were varied.

All spectra were acquired with $T_E = \SI{68}{ms}$ and $T_R = \SI{2000}{ms}$ with $160$ averages and $N=2048$ time samples per spectrum. The sampling frequency was $\SI{1250}{Hz}$ for experiments E1, E2, E3, E4a, and E4c, and $\SI{2000}{Hz}$ for E4b, E4d, E6, E7, E8, E9a, E9b, E11, and E14. These parameters were chosen as they are generally optimal for GABA editing. The sampling frequencies are determined by the dwell time of each measurement. Shorter dwell times allow for faster data acquisition and a broader spectral frequency range, but also reduce the signal-to-noise ratio (SNR). $\SI{2000}{Hz}$ (dwell time $\SI{0.5}{ms}$) is the most common sampling frequency at $\SI{3}{T}$, but $\SI{1250}{Hz}$ (dwell time $\SI{0.8}{ms}$) still covers the frequency range of interest with slightly better SNR, which is why both values were used.

For every MEGA-PRESS run, the acquired time series data was Fourier-transformed, frequency and phase-corrected, and averaged to create three spectra: OFF, ON, and DIFF\@. This was done using a combination of vendor-supplied scanner software and in-house MATLAB code. Acquisition parameters and reconstruction details are described in~\cite{BENCHMARK}.

\begin{table}
  \caption{Composition of the experimental data: $144$ spectra from $112$ phantoms (phantom count is the sum of the first number in each row, ignoring the repetition factor when given as e.g.\ $1 \times 2$ or $8 \times 4$). For E4, $8 \times 4$ denotes 8 gel phantoms each scanned in four acquisition rounds (E4a--E4d). Concentrations are in $\text{mM} = \text{mmol}/\text{L}$. E1 contains a phantom with NAA only and a phantom with NAA and Cr only, followed by phantoms with increasing amounts of GABA (Cr indicated by $0/8.0$ where applicable). E4 phantoms were acquired four times with different bandwidths (E4a/c: $\SI{1250}{Hz}$, E4b/d: $\SI{2000}{Hz}$) and reacquired one week later (E4c/d). E9 consists of $7$ phantoms with varying concentrations of all metabolites; one phantom was scanned three times ($1 \times 3$) and the rest twice ($1 \times 2$), so that E9a contains one more spectrum than E9b.}\label{BenchmarkExperiment}
  \centering
  \resizebox{\textwidth}{!}{%
    \begin{tabular}{llc|ccccp{6.7cm}}
      Series & Medium   & \# Phantoms  & NAA    & Cr      & Glu    & Gln    & GABA                                                                                                                         \\\hline
      E1     & Solution & $13$         & $15.0$ & $0/8.0$ & $0.0$  & $0.0$  & $0.00$, $0.52$, $1.04$, $1.56$, $2.07$, $2.59$, $3.10$, $4.12$, $6.15$, $8.15$, $10.12$, $11.68$                             \\
      E2     & Solution & $1$          & $15.0$ & $0.0$   & $0.0$  & $0.0$  & $0.0$                                                                                                                        \\
      &          & $15$         & $15.0$ & $8.0$   & $0.0$  & $0.0$  & $0.00$, $0.50$, $1.00$, $1.50$, $2.00$, $2.50$, $3.00$, $3.99$, $4.98$, $5.96$, $6.95$, $7.93$, $8.90$, $9.88$, $11.81$      \\
      E3     & Solution & $15$         & $15.0$ & $8.0$   & $12.0$ & $3.0$  & $0.00$, $1.00$, $2.00$, $3.00$, $3.99$, $4.98$, $5.97$, $6.95$, $7.93$, $8.91$, $9.88$, $10.85$, $11.81$, $12.77$, $13.73$   \\
      E4     & Gel      & $8 \times 4$ & $15.0$ & $8.0$   & $12.0$ & $3.0$  & $0.00$, $1.00$, $2.00$, $3.00$, $4.00$, $6.00$, $8.00$, $10.00$                                                              \\
      E6     & Solution & $10$         & $15.0$ & $8.0$   & $12.0$ & $3.0$  & $0.00$, $1.03$, $2.05$, $3.06$, $4.07$, $6.09$, $8.09$, $10.08$, $12.05$, $14.98$                                            \\
      E7     & Solution & $16$         & $12.0$ & $7.0$   & $12.0$ & $3.0$  & $0.00$, $1$, $2$, $2.99$, $3.98$, $4.97$, $5.95$, $6.93$, $7.9$, $8.87$, $9.84$, $10.80$, $11.76$, $12.71$, $13.66$, $14.61$ \\
      E8     & Gel      & $8$          & $15.0$ & $8.0$   & $12.0$ & $3.0$  & $0.00$, $1.00$, $2.00$, $3.00$, $4.00$, $6.00$, $8.00$, $10.00$                                                              \\
      E9     & Gel      & $1 \times 2$ & $14.0$ & $8.0$   & $2.0$  & $11.0$ & $6.0$                                                                                                                        \\
      &          & $1 \times 2$ & $8.0$  & $7.0$   & $7.0$  & $13.0$ & $4.0$                                                                                                                        \\
      &          & $1 \times 2$ & $10.0$ & $6.0$   & $6.0$  & $8.0$  & $6.0$                                                                                                                        \\
      &          & $1 \times 2$ & $12.0$ & $9.0$   & $2.0$  & $14.0$ & $4.0$                                                                                                                        \\
      &          & $1 \times 3$ & $11.0$ & $8.0$   & $3.0$  & $10.0$ & $3.0$                                                                                                                        \\
      &          & $1 \times 2$ & $15.0$ & $10.0$  & $4.0$  & $9.0$  & $5.0$                                                                                                                        \\
      &          & $1 \times 2$ & $13.0$ & $7.0$   & $5.0$  & $12.0$ & $2.0$                                                                                                                        \\
      E11    & Gel      & $8$          & $12.0$ & $7.0$   & $12.0$ & $3.0$  & $0.00$, $1.00$, $2.00$, $3.00$, $4.00$, $6.00$, $8.00$, $10.00$                                                              \\
      E14    & Gel      & $11$         & $11.8$ & $6.3$   & $8.55$ & $2.25$ & $0.00$, $1.02$, $2.05$, $3.06$, $4.08$, $5.09$, $6.1$, $7.1$, $8.1$, $9.1$, $10.09$                                          \\\hline
  \end{tabular}}
\end{table}

\subsection{Preprocessing the Datasets}\label{sec:preproc}

All experimental (phantom) and simulated spectra are processed with a harmonised pipeline so that inputs presented to the models share an identical ppm axis, spectral resolution and scaling. Only the data-origin steps differ; the subsequent export to model inputs is identical across datasets.

For experimental spectra only, the complex free-induction decays (FIDs) for the OFF and ON acquisitions are read. Processing proceeds in the time and then frequency domain:
\begin{enumerate}
  \item Apodization: no windowing is applied, as it did not improve downstream performance in our data (Hamming/Hanning windowing was tried).
  \item Phase handling: when real/imaginary channels are used, we explored applying a linear phase correction (constant and linear terms in frequency) estimated by minimising the imaginary component or spectral entropy of the real part. As no significant improvement was observed, we did not use this for training the models.
  \item Water-peak attenuation: around the water resonance ($\SI{4.75}{ppm} \pm \SI{0.75}{ppm}$ window), large outliers are clamped towards the local median in the magnitude spectrum to reduce residual water without altering phase.
  \item B0 alignment across acquisitions: a single frequency shift (ppm) is estimated per spectrum from a reference metabolite. We examine NAA ($\SI{2.01}{ppm}$) and Cr ($\SI{3.015}{ppm}$) within narrow windows ($\SI{\pm 0.25}{ppm}$) and use Jain's method~\cite{Jain1979} to estimate the exact peak location in the Fourier spectrum. The reference peak with higher prominence is used, and the resulting shift is applied uniformly to OFF, ON and DIFF\@. Residual misalignment after resampling is negligible at the chosen resolution.
  \item Difference spectrum: DIFF is computed as $\text{ON} - \text{OFF}$ after the above steps to ensure consistency with aligned acquisitions.
\end{enumerate}

Simulated spectra required less pre-processing. For consistency with the experimental spectra, we apply B0 alignment in the same way as for the experimental spectra after noise has been added and DIFF is recomputed. No water suppression or additional phase correction is applied.

Independent of their origin, spectra are prepared for training and inference using the same steps:
\begin{enumerate}
  \item Fixed ppm band and resolution: signals are zero-filled/truncated in time and FFT-transformed; the frequency-domain spectra are then oriented to a common descending ppm axis and cropped to \([\SI{4.5}{ppm},\SI{1.0}{ppm}]\) with exactly $2048$ points. This harmonises datasets acquired at different bandwidths (e.g., $\SI{1250}{Hz}$ vs $\SI{2000}{Hz}$) without altering linewidths; zero-filling does not change SNR but provides a common sampling grid. The ppm axis direction is consistent across all inputs.
  \item Amplitude normalisation: each spectrum is scaled by the maximum magnitude of its OFF acquisition when available; otherwise by the global maximum across provided acquisitions. This yields a comparable dynamic range across samples and acquisitions.
  \item Inputs and channels: for each acquisition (OFF, ON, DIFF) we can export real, imaginary and/or magnitude channels, providing up to nine channels; the channel combination and acquisition selection used by the best CNN and YAE configurations are reported in Section~\ref{sec:selection}.
  \item Baseline handling: beyond the water-window attenuation, no explicit baseline modelling or subtraction is applied. In particular, the DIFF baseline is retained so that the models learn to accommodate realistic baselines present in both simulated and experimental data.
  \item Targets (for supervised training): ground-truth values are exported as an ordered vector over the metabolites. Concentrations are expressed relative to NAA as reference, so the NAA entry equals $1$ by construction. During model selection, we considered sum-normalisation (vector divided by its sum) and max-normalisation (divided by its maximum), a common normalisation step to account for variations in total signal or reference metabolite concentrations~\cite{Harris2017}. The normalisation and channel choices for the selected best models are given in Section~\ref{sec:selection}; evaluation metrics in Section~\ref{sec:eval} use max-normalised concentrations.
\end{enumerate}
These steps ensure that spectra presented to the models are aligned in ppm, uniformly sampled, consistently scaled and represented in a common acquisition-channel format across experimental and simulated datasets.

\subsection{Performance Evaluation}\label{sec:eval}

Performance is evaluated in terms of spectral reconstruction (denoising) and metabolite quantification accuracy.

For architectures that output a reconstructed spectrum (our YAE), denoising is assessed by the mean absolute error between the \emph{noise-free} (clean) spectrum and the model's reconstruction from the noisy input:
\begin{equation}
  \epsilon_{\text{spec}} = \frac{1}{F} \sum_{f=1}^{F} |x_f - \hat{x}_f|
\end{equation}
where $x_f$ is the clean reference value at frequency bin $f$, $\hat{x}_f$ is the reconstructed value, and $F$ is the number of frequency points. This quantifies how well the model suppresses noise while preserving spectral structure.

Quantification performance is evaluated using the mean absolute error (MAE) over maximum-normalised concentrations,
\begin{equation}\label{eq:quan-mae}
  \epsilon = \frac{1}{N} \sum_{l=1}^{N} |g_l-p_l|
\end{equation}
where $g_l$ is the ground-truth and $p_l$ the predicted concentration of metabolite $l$, and $N$ is the number of metabolites. We use maximum-normalised concentrations for evaluation regardless of the normalisation used during training, as MAE in this space is robust to outliers and comparable across datasets. Where appropriate, we also report error distributions, mean and standard deviation.

To assess systematic bias and scale, we fit a linear regression of predicted vs.\ ground-truth concentrations, $y = ax + q$. Ideal performance corresponds to slope $a=1$ and intercept $q=0$. We report the slope, intercept, coefficient of determination $R^2$, standard error of the fit, and $p$-value for the regression. Together, MAE and regression metrics characterise both the magnitude of errors and any consistent over- or under-estimation.

\subsection{Experimental Ground-Truth Validation}

The ground truth for the experimental spectra is the millimolar concentrations of the metabolites. Because the algorithms output only normalised concentrations and conversion to millimolar would require additional calibration, we evaluate relative concentrations (ratios eliminate arbitrary scaling). NAA is chosen as a reference for the relative concentrations as it has a prominent peak in both the OFF and DIFF spectra and relatively stable peak characteristics. Alternatively, Cr could be used as a reference, but NAA is preferable here because the Cr signal is absent in the DIFF spectra.

Once the relative concentrations have been calculated, the mean absolute error (MAE) is used to measure the prediction error for each metabolite; when reported over the phantom set, we use the mean absolute error over all spectra (overall MAE) or, for experiment-level comparison, the mean over spectra within each series (Section~\ref{sec:stat_compare}). The standard deviation (SD) is used to reflect the variability of the prediction error. Linear regression fitting and statistical analysis are performed for the relative concentrations in the same manner as detailed above.

Finally, we compare our quantification results with LCModel, one of the most widely used tools for \emph{in vivo} MRS quantification. LCModel was run separately on the OFF and DIFF spectra of each phantom. For a fair comparison with the DL models, LCModel was configured with a basis set derived from the same FID-A simulations and Hamiltonian choices used to generate the DL training data (Section~\ref{sec:data_sim}), including the Near model for GABA\@. Fitting range, baseline model (spline), and other settings were chosen to match typical MEGA-PRESS practice; further details (version, fitting range, baseline parameters) are given in the benchmark study~\cite{BENCHMARK}. Table~\ref{tab:lcmodel_off_vs_diff} compares quantification accuracy of LCModel-OFF and LCModel-DIFF across experiments; LCModel-OFF achieved lower experiment-level errors for GABA and Glu, and is therefore used as the main conventional baseline in the following comparisons. We use LCModel-OFF as the conventional baseline because it gave lower errors than LCModel-DIFF on our phantom data (Table~\ref{tab:lcmodel_off_vs_diff}), so the DL models are compared against a strong rather than a weak baseline; this aligns with literature suggesting off-resonance spectra can yield more reliable quantification than difference spectra under certain conditions~\cite{LCModelManual}.

\subsection{Statistical Comparison of Quantification Errors}\label{sec:stat_compare}

To compare the quantification accuracy between models on phantom data in a statistically principled and model-agnostic manner, we further analyse the distributions of absolute quantification errors across experiments, where a single experiment here refers to one series in the phantom data. Since absolute errors are non-negative and typically non-Gaussian, all statistical analyses are based on non-parametric methods.

To avoid pseudo-replication arising from treating spectra acquired under identical experimental conditions as independent samples, each experiment is treated as the fundamental statistical unit. Thus, experiment-level MAE denotes the mean error over spectra within each series (one value per series per metabolite), whereas overall MAE denotes the mean error over all $144$ spectra as used later in summary Tables~\ref{tab:overall_mae} and~\ref{tab:linewidth_aug_mae}. For a given experiment $e$ and metabolite $m$, the mean absolute error across spectra is computed as
\begin{equation}
  \tilde{\epsilon}_{e,m} = \frac{1}{K_e} \sum_{k=1}^{K_e} \epsilon_{e,k,m},
\end{equation}
where $K_e$ denotes the number of spectra in experiment $e$. The mean provides a standard summary of typical error magnitude per experiment.

Paired one-tailed Wilcoxon signed-rank tests are then applied to the experiment-level aggregated errors to assess whether one model consistently yields lower quantification errors than another across experiments. This paired formulation exploits the fact that both models are evaluated under identical experimental conditions, thereby isolating differences attributable to the quantification method.

In addition to hypothesis testing, effect sizes are reported to facilitate direct interpretation of performance differences. Specifically, the proportion of experiments with lower error,
\begin{equation}
  P_e = \frac{1}{E} \sum_{e=1}^{E} \mathbb{I}\!\left(\tilde{\epsilon}^{(1)}_{e,m} < \tilde{\epsilon}^{(2)}_{e,m}\right),
\end{equation}
quantifies the fraction of experiments in which one model outperforms the other, while the mean difference in absolute error,
\begin{equation}
  \Delta_m = \frac{1}{E} \sum_{e=1}^{E} \left(\tilde{\epsilon}^{(1)}_{e,m} - \tilde{\epsilon}^{(2)}_{e,m}\right),
\end{equation}
indicates the direction and magnitude of the performance difference. We use these metrics to compare models consistently: MAE and regression slope/intercept for magnitude and bias, and experiment-level Wilcoxon tests and effect sizes for statistical comparison.

\section{Deep Learning Architectures}\label{sec:archs}

This section describes the deep learning architectures evaluated here. We first present our two proposed architectures---a convolutional multi-class regressor (CNN) and a Y-shaped autoencoder (YAE)---which are extensively parameterised and optimised via Bayesian model selection (Section~\ref{sec:selection}). We then describe several comparative baselines from the literature, implemented with their published configurations and minimal adaptations to our MEGA-PRESS pipeline.

The \textbf{CNN} builds on the architecture of~\cite{MRSNET} and consists of 1D and 2D convolutions that extract features along the frequency axis and across acquisition channels, followed by fully connected layers that regress to metabolite concentrations. We explore a broader architectural space (kernel sizes, down-sampling, regularisation, activations) to identify an optimal configuration.

The \textbf{YAE} is a two-branch network: an encoder maps the input to a latent representation; one branch (decoder) reconstructs a denoised spectrum, and the other (quantifier) regresses the latent representation to concentrations. This design is motivated by the physical principle that an MRS spectrum is a linear combination of basis spectra: the autoencoder encourages a latent space that captures the weights of these basis components, which the quantifier then maps to concentrations. The decoder acts as a regulariser, ensuring the latent representation retains sufficient spectral information for reconstruction.

For these architectures, the input can be chosen from ON, OFF, and DIFF acquisitions and from real, imaginary, or magnitude representations; see Section~\ref{sec:preproc}. The following subsections detail the CNN (Section~\ref{sec:CNN}), the YAE (Section~\ref{sec:archs_yae}), and the additional baselines (Section~\ref{sec:additional_baselines}).

\begin{table}
  \caption{Parameterised convolutional multi-class regressor (CNN) architecture with parameters. $N_f$ is the number of frequencies in the inputs, and channels refers to the number of input spectra. The parameters $f_1$ and $f_2$ determine the number of convolutional filters, $k_1$, $k_2$, $k_3$ and $k_4$ determine the kernel size of the convolutions, $s_1$ and $s_2$ determine the use of strides or max-pooling, $d_1$ determines the use of dropout or batch normalisation, $d_2$ the dropout rate in the dense layer, and $e$ the number of neurons in the dense layer. The activation function at the output is either sigmoid or softmax, and the output consists of $N_m$ metabolite concentrations (fixed to $5$).}\label{tab:cnn}
  \resizebox{\textwidth}{!}{%
    \begin{tabular}[t]{lp{.36\textwidth}}\hline
      Layer      & Description                                                                 \\\hline
      input      & (channels, $N_f$)                                                           \\\hline
      conv1      & FCONV ($f_1$, $(1,k_1)$, $s_1$, $d_1$)                                      \\
      conv2      & FCONV ($f_1$, $(1,k_2)$, $s_1$, $d_1$)                                      \\\hline
      reduce     & Repeat until one output channel:                                            \\
      & \hspace*{0.6em}FCONV($f_1$, ($\min(\text{channels},3)$, $k_3$), $1$, $d_1$) \\\hline
      conv3      & FCONV ($f_1$, $(1,k_4)$, $1$, $d_1$)                                        \\
      conv4      & FCONV ($f_1$, $(1,k_4)$, $s_2$, $d_1$)                                      \\
      conv5      & FCONV ($f_2$, $(1,k_4)$, $1$, $d_1$)                                        \\
      conv6      & FCONV ($f_2$, $(1,k_4)$, $s_2$, $d_1$)                                      \\\hline
      dense1     & sigmoid (Dense ($e$))                                                       \\
      dropout    & if $d_2 > 0.0$ then Dropout ($d_2$)                                         \\
      dense2     & Dense (metabolites)                                                         \\
      activation & sigmoid () or softmax ()                                                    \\\hline
      output     & $N_m$                                                                       \\\hline
    \end{tabular}\quad\hfill
    \parbox[t]{.49\textwidth}{\vspace*{1ex}
      FCONV ($f$,$k$,$s$,$d$) is a sequence of layers forming mainly a convolution, determined by its parameters as follows:\\
      {\tt
        if $s \leq 0$:\\
        \hspace*{1.2em} $x$ = Conv2D (filter=$f$, kernel\_shape=$k$)($x$)\\
        else:\\
        \hspace*{1.2em} $x$ = Conv2D (filter=$f$, kernel\_shape=$k$,\\
        \hspace*{7.25em} strides = (1,$s$)) ($x$)\\
        if $d == 0.0$:\\
        \hspace*{1.2em} $x$ = BatchNormalisation () ($x$)\\
        $x$ = ReLU () ($x$)\\
        if $d > 0.0$:\\
        \hspace*{1.2em} $x$ = Dropout ($d$) ($x$)\\
        if $s < 0$:\\
        \hspace*{1.2em} $x$ = MaxPool2D ($(1,|s|)$) ($x$)
  }}}
\end{table}

\subsection{CNN\@: Convolutional Multi-Class Regressor}\label{sec:CNN}

Our parameterised CNN architecture is shown in Table~\ref{tab:cnn}. It consists of a sequence of convolutional layers followed by two dense layers. Its input has a channel axis, for the different acquisitions and datatypes, followed by a frequency axis for the bins from the Fourier transform; see Section~\ref{sec:preproc}.

The convolutional blocks are represented by FCONV ($f$, $k$, $s$, $d$), a sequence of layers detailed in Table~\ref{tab:cnn}. This block primarily consists of a convolution with $f$ filters, kernel size $k$, and a ReLU activation function. The parameter $s$ controls dimensionality reduction along the frequency axis, using either strides (if $s > 0$) or max-pooling (if $s < 0$). The parameter $d$ controls regularisation: dropout is used if $d > 0$, and batch normalisation is used if $d=0$. This parameterisation enables us to explore different strategies for down-sampling and regularisation during model selection. While batch normalisation addresses internal covariate shift to improve training, dropout regularisation addresses overfitting by learning more robust features. In practice, using both simultaneously is often redundant.

The first two layers are 1D convolutions along the frequency axis (kernel sizes $k_1$, $k_2$), applied per channel to detect local spectral features. A reduction module then collapses the channel dimension to one via sequential 2D convolutions with kernel size $(\min(\text{channels},3),k_3)$, combining information across acquisitions; this channel-then-reduce design follows~\cite{MRSNET}. A further sequence of 1D convolutions along the frequency axis on this single combined channel extracts higher-level spectral features. Two dense layers (with sigmoid activation between them and optional dropout) map these features to concentrations via a final softmax or sigmoid output. While softmax may match the relative concentrations better, sigmoid may be more suitable as the relative concentrations do not represent probabilities. The network is trained by minimising the mean squared error (MSE). Training and model selection are detailed in Section~\ref{sec:selection}.

\subsection{Fully Connected Y-Shaped Autoencoder (YAE)}\label{sec:archs_yae}

\begin{figure}
  \centering
  \includegraphics[width=1.0\textwidth]{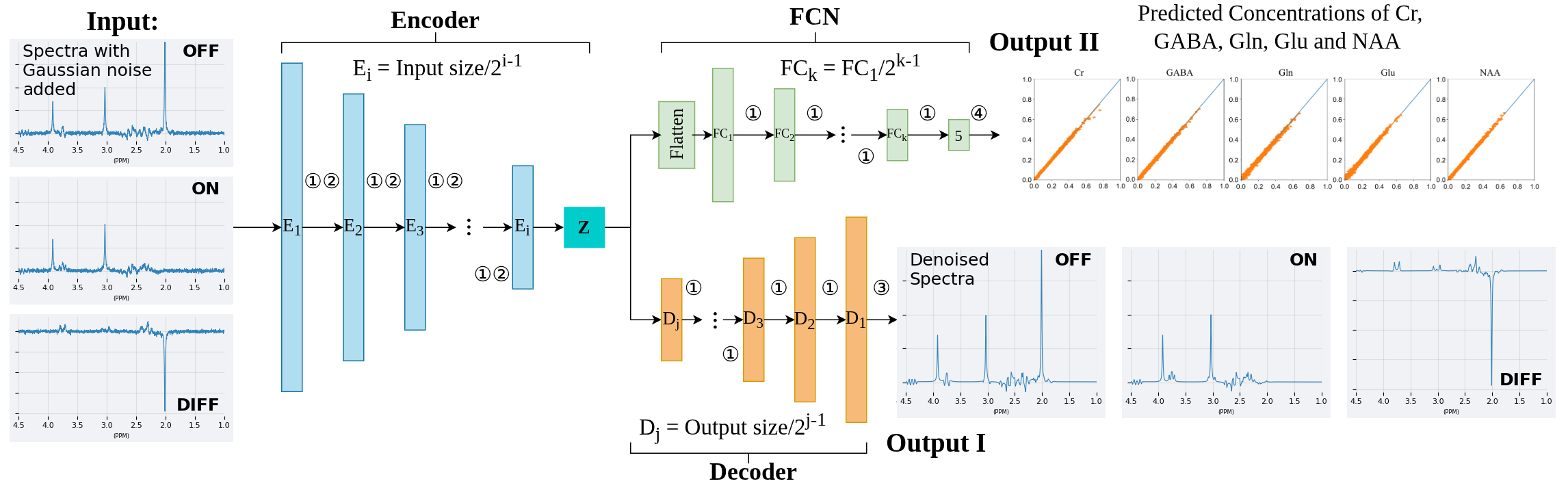}
  \caption{Illustration of the Y-shaped autoencoder (YAE) architecture. The input consists of one or more noisy MEGA-PRESS spectra (OFF, ON, DIFF using real, imaginary or magnitude representations). The encoder maps the input to a compressed latent representation. The decoder branch reconstructs denoised versions of the input spectra from this latent space. The quantifier branch predicts the metabolite concentrations from the same latent representation. Key components are highlighted: \ding{172} hidden layer activation function, \ding{173} dropout layer, \ding{174} decoder output activation function, and \ding{175} quantifier output activation function.}\label{fig:AE-MRSNet}
\end{figure}

The YAE is a deterministic, fully connected network with three modules: an encoder, a decoder, and a quantifier (Figure~\ref{fig:AE-MRSNet}, Table~\ref{tab:yae}). We use fully connected (rather than convolutional) layers to capture global spectral structure: in the frequency domain, information lies in overall lineshapes rather than local patterns, and CNNs' locality bias can be misaligned with this. We do not use a variational autoencoder (VAE) design because stochastic latents and VAE regularisation can reduce reconstruction fidelity, which we need for precise quantification. Input and output options match the CNN (Section~\ref{sec:preproc}): the decoder outputs a denoised spectrum; the quantifier outputs normalised relative concentrations.

\begin{table}
  \caption{Parameterised architecture for the YAE's encoder, decoder, and quantifier modules. The specific values explored for the parameters are given in Section~\ref{sec:selection}. Channels refer to the number of channels used as input spectra. $N_f$: number of frequency samples in the input spectrum. $N_q$: number of neurons in the first dense layer of the quantifier. $N_m$: number of output metabolite concentrations (fixed to $5$). $L_e, L_d, L_q$: number of dense layers in the encoder, decoder, and quantifier, respectively. $a_e, a_d$: activation function for the encoder and decoder. $a_q, a_m$: activation function for the hidden layers and output layer of the quantifier. $d_e$: dropout rate for the encoder.}\label{tab:yae}
  \begin{tabular}[t]{lp{.19\textwidth}}
    \multicolumn{2}{c}{\bf Encoder}                                                                                      \\\hline
    Layer                    & Description                                                                               \\\hline
    input                    & \rule[-1.5ex]{0pt}{4.2ex}$(\text{channels}, N_f)$                                         \\\hline
    dense\_e                 & for each channel c:                                              \\
    & \hspace*{0.6em} for $l = 1, \dots, L_e -1$:                                               \\
    & \hspace*{1.2em} Dense$\left(\tfrac{N_f}{2^{l-1}}, a_e\right)$                             \\
    & \hspace*{1.2em} Dropout$\left(d_e\right)$                                                 \\
    \hspace*{0.6em}latent\_c & \rule[-1.5ex]{0pt}{4.2ex}\hspace*{0.6em} Dense$\left(\tfrac{N_f}{2^{L_e -1}}, a_e\right)$ \\\hline
    output                   & \rule[-1.5ex]{0pt}{4.2ex}$\left(\frac{N_f}{2^{L_e -1}}\right)$ per channel                \\\hline
  \end{tabular}\hfill
  \begin{tabular}[t]{lp{.19\textwidth}}
    \multicolumn{2}{c}{\bf Decoder}                                                                        \\\hline
    Layer                    & Description                                                                 \\\hline
    input                    & \rule[-1.5ex]{0pt}{4.2ex}$\left(\tfrac{N_f}{2^{L_e -1}}\right)$ per channel \\\hline
    dense\_d                 & for each channel c:                                                         \\
    & \hspace{0.6em}for $l = L_d, \dots, 2$:                                      \\
    & \hspace*{1.2em} Dense$\left(\tfrac{N_f}{2^{l-1}}, a_d\right)$               \\\\
    \hspace*{0.6em}decode\_c & \rule[-1.5ex]{0pt}{4.2ex}\hspace{0.6em}Dense$\left(N_f, a_d\right)$         \\\hline
    output                   & \rule[-1.5ex]{0pt}{4.2ex}${(N_f)}$ per channel                              \\\hline
  \end{tabular}\hfill
  \begin{tabular}[t]{lp{.19\textwidth}}
    \multicolumn{2}{c}{\bf Quantifier}                                                    \\\hline
    Layer    & Description                                                                \\\hline
    input    & \rule[-1.5ex]{0pt}{4.2ex}$\left(\frac{N_f}{2^{L_e -1}}\right)$ per channel \\\hline
    flatten  & $\text{channels} \times \text{frequencies}$                                \\[-.5pt]\hline
    dense\_q & for $l = 1, 2, \dots, L_q-1$:                                              \\
    & \hspace*{0.6em} Dense$\left(\frac{N_q}{2^{l-1}}, a_q\right)$                        \\
    \\\hline
    quant    & \rule[-1.5ex]{0pt}{4.2ex}Dense$\left(N_m, a_m\right)$                      \\\hline
    output   & \rule[-1.5ex]{0pt}{4.2ex}$N_m$                                             \\\hline
  \end{tabular}\hfill\rule{0pt}{1ex}
\end{table}

For each channel of the input, the encoder consists of a sequence of fully connected layers, reducing the input frequency dimensions while enriching the feature representation in the latent space. As detailed in Table~\ref{tab:yae}, each layer comprises $\tfrac{N_f}{2^{l-1}}$ neurons at the $l$-th layer, where $N_f$ denotes the initial frequency dimension. Dropout layers ($d_e$) are applied after each dense layer for regularisation, so the model can be adapted to different complexities and dataset sizes. The final encoder layer compresses the input into a low-dimensional latent representation of size $\tfrac{N_f}{2^{L_e - 1}}$ per channel. The latent representation feeds into two separate branches, forming the ``Y'' shape.

The first branch is the decoder, which aims to reconstruct a denoised spectrum from the latent representation, restoring the original input shape of $(\text{channels}, N_f)$. Its architecture consists of a sequence of fully connected layers that progressively expand the data, structured to reverse the encoder's compression (Table~\ref{tab:yae}). The decoder is used to ensure the latent representation contains sufficient features to reconstruct a clean spectrum, rather than for denoising. This is motivated by the physical principle that an MRS spectrum is a linear combination of basis spectra (see Section~\ref{sec:data_sim}) and the demonstrated efficacy of denoising autoencoders in MRS~\cite{LAM,YangLei}. Dropout was not applied in the decoder module, as its objective is to reconstruct the spectrum in a stable and deterministic manner; stochastic regularisation here could disrupt reconstruction fidelity.

The second branch is the quantifier, which regresses the latent representation to metabolite concentrations. The latent representation is flattened across channels and frequency dimensions, then passed through a sequence of fully connected layers ($\tfrac{N_q}{2^{l-1}}$ neurons at layer $l$). We do not use dropout in the quantifier: the encoder already regularises the latent space, and dropout at the final regression stage would add unnecessary stochasticity where a stable, deterministic mapping to concentrations is required. The decoder acts as a regulariser, as the encoder must retain enough spectral information such that the decoder can reconstruct the full spectrum. So the latent representation cannot collapse to quantification-only features. As an MRS spectrum is a linear combination of basis spectra (Section~\ref{sec:data_sim}), this aims to ensure the latent space reflects those weights. The final output layer of the quantifier maps the transformed features to $N_m$ target dimensions: the number of quantified metabolites. The Y-shape ties the latent representation to both reconstruction and quantification, so the encoder is encouraged to learn features that support both tasks rather than overfitting to concentration targets alone.

We use the Huber loss for the decoder and quantifier branches:
\begin{equation}
  S_{\delta}(x,x') =
  \begin{cases}
    \frac{1}{2} {(x - x')}^{2},                      & \text{if } \left|x-x'\right| \le \delta, \\
    \delta (\left|x-x'\right| - \frac{1}{2} \delta), & \text{otherwise.}
  \end{cases}
\end{equation}
where $x$ is the ground truth, $x'$ is the prediction, and $\delta$ is a hyperparameter. We use $\delta=1.0$, the default value in TensorFlow. When the absolute error is less than or equal to $\delta$, the Huber loss is quadratic like MSE\@; for larger errors, it becomes linear like MAE\@. This makes the loss less sensitive to large errors and outliers compared to MSE\@.

This choice is motivated by the nature of MRS data. For the decoder branch, the loss $S_{\delta}(y,y')$ is calculated between the noise-free ground truth spectrum $y$ and the reconstructed spectrum $y'$. In MEGA-PRESS spectra, some metabolite signals are very prominent and could dominate an MSE loss. Huber loss mitigates the influence of these large signals and helps the model reconstruct the finer details of less prominent spectral features, such as those from GABA\@. For similar reasons, we also use the Huber loss $S_{\delta}(g,p)$ for the quantifier branch, where $g$ is the ground truth and $p$ the predicted vector of concentrations.

The specific methodology used to select the optimal parameters for this architecture, along with the training strategy, is detailed in Section~\ref{sec:selection}.

\subsection{Additional Deep Learning Baselines}\label{sec:additional_baselines}

We implement four further DL models from the literature as comparative baselines. These are used with their \emph{published default configurations} with minimal adaptations needed to handle the MEGA-PRESS input and train on the quantification objective. They are not subject to the Bayesian optimisation applied to the CNN and YAE, due to computational constraints; our results therefore reflect their out-of-the-box performance in this setting rather than a fully tuned comparison. Pipeline adaptations may also affect comparability with the originally reported results.

\textbf{FCNN}~\cite{Hatami2018} is a seven-layer 1D-CNN using five 1D convolutional layers with concatenated ReLU (CReLU) activation, followed by a two-layer dense head for regression. We adopted the published architectural parameters, including the convolutional filter progression ($[32, 64, 128, 256, 512]$) and $1024$ units in the first dense layer. The original model was designed for two-channel (real, imaginary) input; our implementation reshapes the arbitrary input acquisitions and datatypes (e.g., OFF, ON, DIFF) to a two-channel or four-channel format to match the model's expected input shape.

\textbf{QNet}~\cite{QNet2024} is a physics-informed model with an IF (Imperfection Factor) Extraction Module (three-block CNN) and an MM (Macromolecule) Signal Prediction Module; concentrations are obtained via a Linear Least Squares (LLS) solver. We adapt it for multi-acquisition data by running separate IF modules per acquisition (e.g., OFF, DIFF) and concatenating their outputs. To focus solely on quantification and due to the absence of macromolecule signal, we disable the MM module and train only the quantification path. Two LLS variants are evaluated: \textbf{QNet} (simplified, learnable BasicLLSModule) and \textbf{QNetBasis} (physics-grounded BasisLLSModule that applies the predicted IFs to a known metabolite basis set). Published defaults are used for the IF module (e.g., $[16, 32, 64]$ filters).

\textbf{QMRS}~\cite{Q-MRS2024} combines a backbone of CNNs, Inception modules, and a Bidirectional LSTM (BiLSTM) with a Multi-Head MLP designed to predict various spectral parameters. We feed a two-channel input (adapted from our pipeline) and train only the metabolite amplitude (concentration) head; other heads (phase, linewidth, baseline) are disabled to focus on quantification and avoid gradient issues. Default parameters are used ($32$ initial filters, $128$ LSTM units, $[512, 256]$ MLP units).

\textbf{EncDec}~\cite{encdec2023}, an encoder-decoder network, uses a sequence of WaveNet blocks for feature extraction, which are then integrated across acquisitions using an attention-based GRU (AttentionGRU). It was designed for JPRESS data, which consists of many echoes (e.g., $32$). We added a custom ContextConverter layer that reshapes the pipeline output from (batch, acquisitions$\times$datatypes, freqs) to (batch, acquisitions, freqs, datatypes) for the EncDec backbone. Similar to QMRS, the auxiliary heads for predicting FIDs and phase parameters were disabled during training to focus solely on quantification. Published parameters are used (e.g., $128$ filters for WaveNet blocks).

\section{Model Selection and Performance on Simulated Data}\label{sec:selection}

We optimise hyperparameters for the CNN and the YAE using similar protocols. We first describe the optimisation strategy, then present model selection for the CNN (including validation that Bayesian optimisation is likely to find the grid-search optimum), then the three-stage YAE selection, and finally the performance of the selected models on simulated data. The process explores various combinations of model hyperparameters, input data representations (real, imaginary, magnitude of ON, OFF, DIFF spectra) and concentration normalisation strategies (sum vs.\ max normalisation; see Section~\ref{sec:preproc}). Both models are trained with the ADAM optimiser ($\beta_1=0.9$, $\beta_2=0.999$, $\epsilon=10^{-8}$) with random batch shuffling. A fixed learning rate of $10^{-4} \times (\text{batch\_size} / 16)$ is used, which was determined to give good convergence in preliminary experiments and reflects a linear scaling rule with batch size~\cite{Goyal2017}.

\subsection{Hyperparameter Optimisation Strategy}

For this multi-dimensional optimisation problem, we employ Bayesian optimisation with a Gaussian process, a sample-efficient method well-suited for tuning expensive black-box functions like deep learning model training~\cite{snoek2012}. As each configuration is tested, the Gaussian process is updated to better represent the performance landscape. We alternate between choosing a configuration based on the Expected Improvement (EI) and selecting configurations by Thompson sampling. EI employs an L-BFGS optimiser to find the maximum of the expected improvement. Thompson sampling takes a possible version of the estimated average MAE function, sampled from the Gaussian process, and acts as if that version were the truth, choosing the best point based on that sample. Interleaving the Thompson with the EI strategy explored the configuration space in a more balanced manner. The iterative process continues until a predetermined number of iterations is reached, and then selects the best model based on the estimated average MAE across the five-fold cross-validation. For each configuration proposed by the optimiser, the corresponding model is trained and evaluated using five-fold cross-validation on a dataset of $10{,}000$ simulated spectra. The mean absolute error (MAE) on the validation folds serves as the objective function to be minimised. Our implementation uses GPyOpt~\cite{gpyopt2016}.

\subsection{CNN Model Selection}\label{sec:selection_cnn}

We run a grid search over the CNN parameter space, then apply Bayesian optimisation on the same space to confirm it recovers the same optimum and to quantify its efficiency. The parameters and their options are simplified, based on the original values explored in~\cite{MRSNET}. Broader parameter exploration did not yield significantly improved models, and is not further explored here for simplicity. The full results are available in the CNN models data repository~\cite{repo-model-cnn}.

\begin{table}
  \caption{Grid search model selection parameters for CNN model, varying the dataset options, exploring different convolutional kernel sizes (small, medium, large), activation functions and batch sizes, leaving the remaining model parameters fixed.}\label{tab:cnn-simple-all}
  \centering
  \begin{tabular}{llll}
    \textbf{Group} & \textbf{Parameter} & \textbf{Values}                                                 & \textbf{Best Model} \\\hline
    Dataset        & Normalisation      & sum, max                                                        & sum                 \\
    & Acquisitions       & (DIFF, OFF), (DIFF, ON), (OFF, ON), (DIFF,OFF,ON)               & (OFF,ON)            \\
    & Datatypes          & (magnitude), (real), (imaginary, real)                          & (real)              \\\hline
    Model          & Kernel sizes       & small:  $k_1=7,  k_2=5, k_3=3, k_4=3$                           &                     \\
    &                    & medium: $k_1=9,  k_2=7, k_3=5, k_4=3$                           & medium              \\
    &                    & large:  $k_1=11, k_2=9, k_3=7, k_4=3$                           &                     \\
    & Activations        & softmax: softmax with $s_1 = 2, s_2=3$                          &                     \\
    &                    & sigmoid-pool: sigmoid with $s_1=-2, s_2=-3$                     & sigmoid-pool        \\
    & Fixed parameters   & $f_1=256, f_2=512, d_1=0, d_2=0.3, e=1024$, $N_f=2048$, $N_m=5$ &                     \\\hline
    Training       & batch size         & $16$, $32$, $64$                                                & $16$                \\\hline
  \end{tabular}
\end{table}

\begin{figure}
  \centering
  \includegraphics[width=\textwidth]{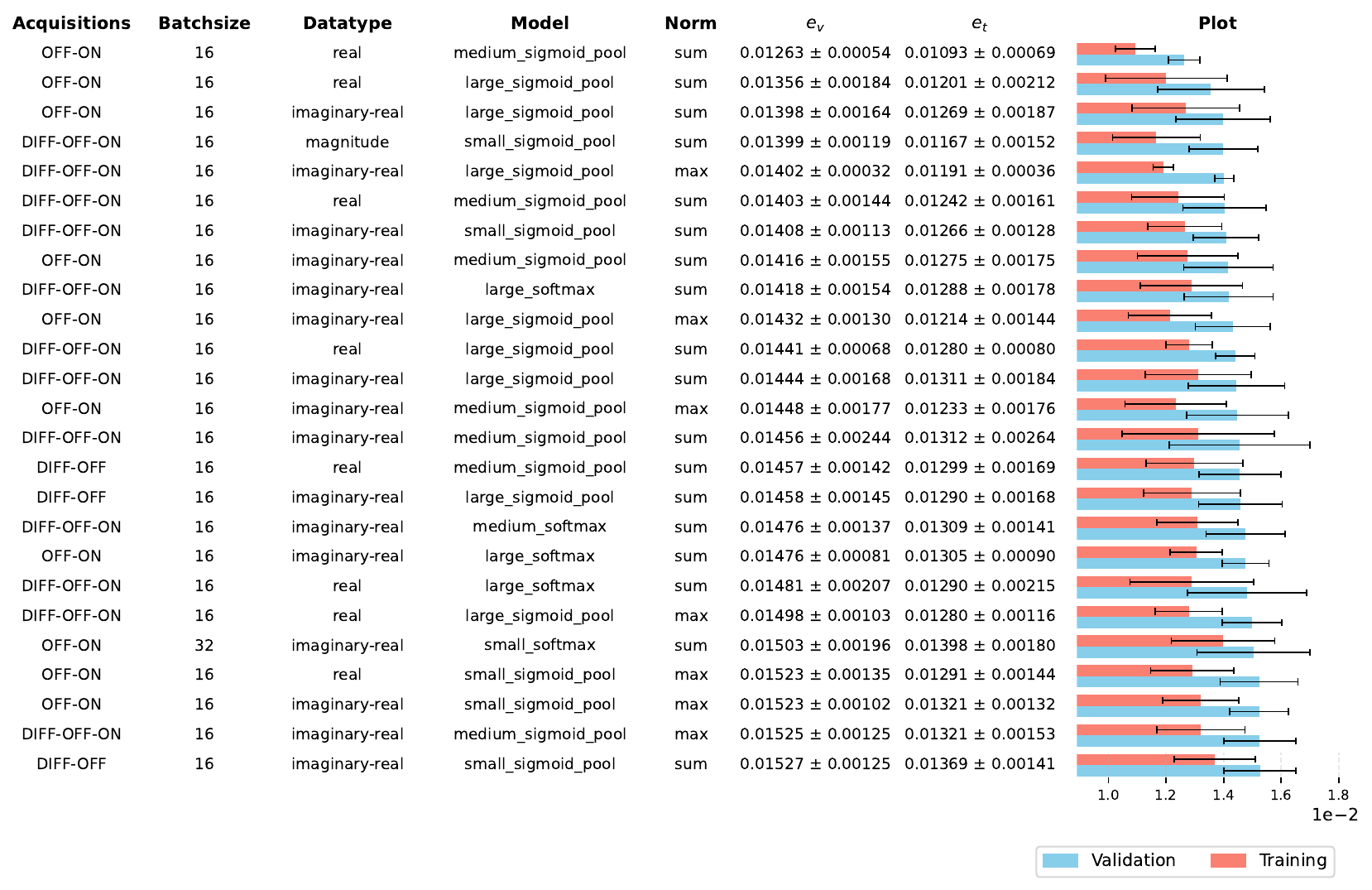}
  \caption{Validation and training concentration MAE for the top $25$ of $432$ CNN configurations from the grid search model selection (Table~\ref{tab:cnn-simple-all}). Each pair of horizontal bars corresponds to one configuration: blue, validation MAE\@; red, training MAE\@. Error bars show standard deviation across five-fold cross-validation. Configurations are ordered by validation MAE (best at top). Dataset: $10{,}000$ simulated spectra, sum normalisation, basis set linewidth $\SI{2}{Hz}$, $100$ epochs.}\label{fig:cnn-simple-all-sel}
\end{figure}

The model parameters for the grid search are listed in Table~\ref{tab:cnn-simple-all}. The options explore the sum and maximum normalisation for the relative concentration output, and different combinations of ON, OFF, DIFF acquisitions, and datatype. For the model parameters, different kernel sizes (small, medium and large) and two options are explored for the final activation functions: softmax with strides or sigmoid with max-pooling. The other model parameters are fixed. We also explore different batch sizes. See Table~\ref{tab:cnn} for details of how the parameters are used by the CNN model architecture.

The results for the top $25$ performing models from the grid search are shown in Figure~\ref{fig:cnn-simple-all-sel}. Analysis of the full grid search results reveals several clear trends. Sum normalisation of target concentrations consistently outperforms maximum normalisation. Models using the real part of the spectra as input, either alone or with the imaginary part, tend to perform better than those using magnitude data. The choice of acquisitions is less clear, though combinations including OFF and ON spectra are consistently among the top models. For the architecture itself, using max-pooling with a sigmoid output activation (sigmoid-pool) is superior to using strides with a softmax output.

Overall, the best configuration identified by the grid search is a model with the medium kernel sizes and sigmoid-pool activation, trained on the real part of the OFF and ON spectra with sum normalisation and a batch size of $16$. While there is a clear distinction between this model and lower-performing ones, some groups exhibit similar performance but still fall short of this optimal model. Large kernel sizes may be worth considering for larger training datasets with more variation, but their higher computational cost is not justified in this context. This optimal configuration differs from that in preliminary work~\cite{MRSNET}; we attribute the difference to the broader search used here, which explored more of the configuration space.

\begin{figure}
  \centering
  \includegraphics[width=.6\textwidth]{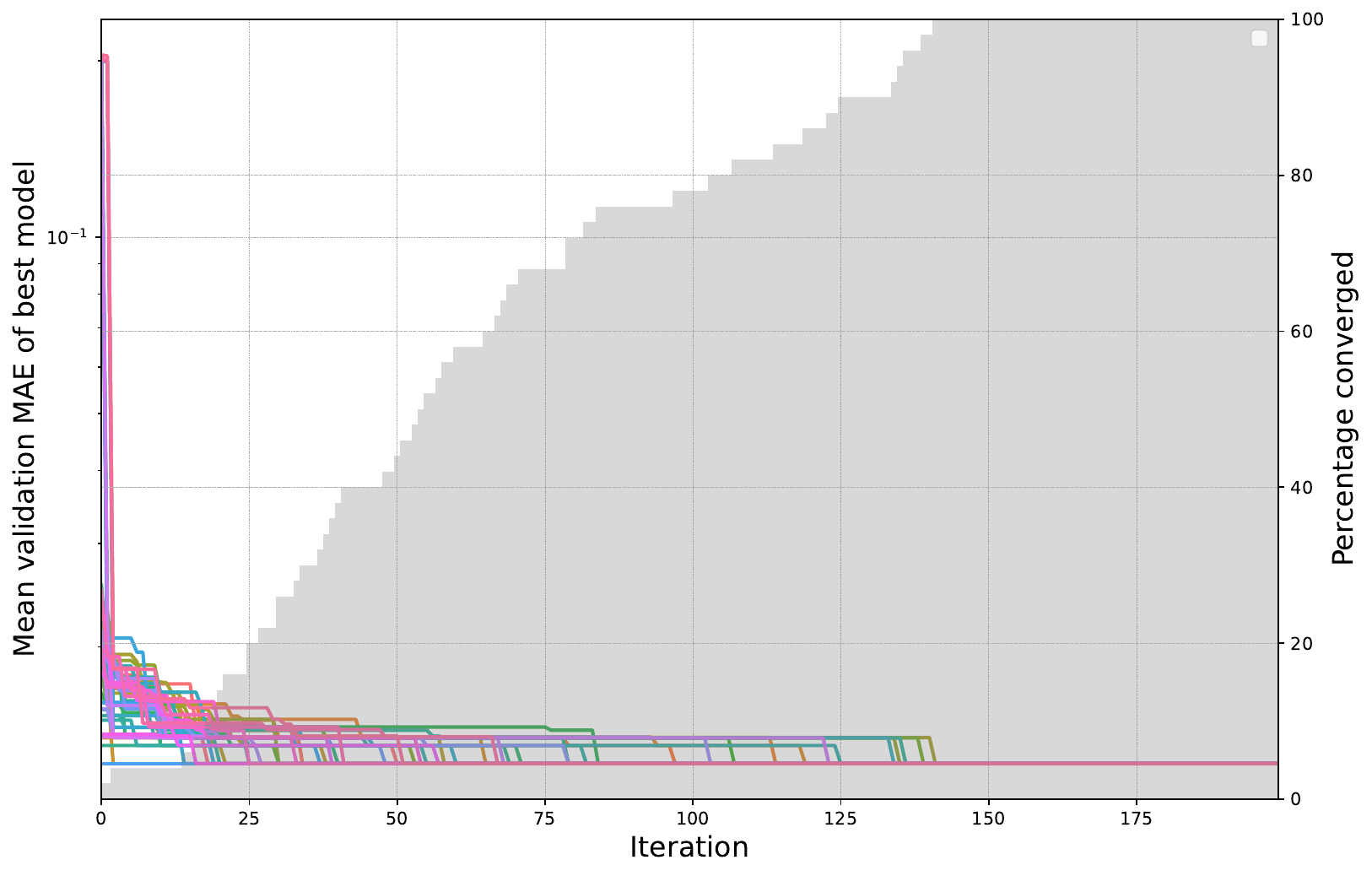}
  \caption{Performance of $50$ repetitions of the Bayesian optimisation for the CNN model selection over the Bayesian optimisation iterations. The grey shaded area (right axis) shows the percentage of runs that selected the same best configuration as the full grid search (``converged'').}\label{fig:cnn-gpo-iters}
\end{figure}

To verify the efficiency of the Bayesian optimisation model selection compared to the grid search, we ran it over the same model parameters listed in Table~\ref{tab:cnn-simple-all}. Repeated runs of the Bayesian optimisation over the CNN model configurations, trained over $100$ epochs, consistently gave the same best model, also found by the grid search over $432$ models. Figure~\ref{fig:cnn-gpo-iters} shows the performance of repeated Bayesian optimisation over the iterations. The best model was found before $150$ models have been explored, about $35\%$ of the model configuration space. This gives us some confidence that the process works and can also be applied to the YAE architecture.

\subsection{YAE Model Selection}\label{sec:selection_yae}

As the parameter space for the YAE architecture is substantially larger than that of the CNN, an exhaustive grid search is computationally infeasible. We therefore rely on the Bayesian optimisation strategy validated in Section~\ref{sec:selection_cnn}. However, the combined parameter space of the full, jointly-trained YAE architecture (described in Section~\ref{sec:archs_yae}) is still too large for a direct optimisation. To keep the search computationally feasible we used an incremental strategy in three stages:
\begin{enumerate}
  \item \textbf{Stage~1: Denoising Autoencoder Optimisation.} To find an optimal architecture for feature extraction, we focused solely on the autoencoder's denoising capability. We performed a parameter search to identify the most effective encoder ($L_e$) and decoder ($L_d$) architectures, along with optimal activation functions and regularisation strategies, for the reconstruction task.
  \item \textbf{Stage~2: Exploratory Quantifier Search (Frozen Encoder).}  We conducted a preliminary Bayesian optimisation on the quantifier module. For this exploratory stage, we attached a quantifier to the fixed and frozen optimal encoder from Stage~1. The goal of this stage was not to find the final model, but rather to efficiently identify the optimal architecture (e.g., $N_q, L_q$) and parameter ranges for the quantifier. The results of this intermediate step were used to prune the search space for the final optimisation stage.
  \item \textbf{Stage~3: Final Joint-Training Optimisation.} Finally, using the optimal architectures and refined parameter ranges from Stages~1 and~2 as a highly-informed starting point, we conducted the definitive model selection on the full, unfrozen, end-to-end joint-training model. This step fine-tuned all parameters simultaneously, leveraging the loss-weighting and ramp-up strategy. The results of this final stage are presented below.
\end{enumerate}

The options for data types, acquisition combinations, and dataset used in the model selection process are kept consistent with those for the CNN model selection (except for using $200$ training epochs, versus $100$ for the less complex CNN), ensuring comparability.

\subsubsection{Structural Exploration and Search Space Pruning (Stage~1 \& 2)}\label{sec:structural_exploration}

To effectively navigate the vast hyperparameter space of the YAE architecture, we conducted extensive exploratory experiments in the first two stages. The goal was not to identify a single ``final'' model, but to empirically determine the impact of key architectural components and prune the search space for the final joint optimisation.

\begin{table}
  \caption{Model selection parameters for the YAE encoder-decoder path.}\label{tab:ae-para}
  \centering
  \begin{tabular}{lll}
    \textbf{Group} & \textbf{Parameter} & \textbf{Values}                      \\ \hline
    Dataset        & Normalisation      & sum                                  \\
    & Acquisitions       & (DIFF, OFF, ON), (DIFF, ON), (DIFF, OFF), (OFF, ON) \\
    & Datatypes          & (real), (imaginary), (magnitude)                    \\ \hline
    Model          & $L_e$              & $5, 6, 7, 8, 9$                      \\
    & $L_d$              & $5, 6, 7, 8, 9$                                     \\
    & $a_e$              & tanh, ReLU (magnitude)                              \\
    & $a_d$              & tanh, sigmoid (magnitude)                           \\
    & Fixed Parameters   & $d_e=0.3$, $N_f=2048$, $N_m=5$                      \\ \hline
    Training       & batch size         & $16, 32, 64$                         \\ \hline
  \end{tabular}
\end{table}

\textbf{Stage~1: Robustness Analysis of the Encoder-Decoder Backbone}
In the first stage, approximately $900$ independent autoencoder models were trained to optimise the self-supervised reconstruction task for the search space of Table~\ref{tab:ae-para}. The full selection results can be found in the data repositories~\cite{repo-model-ae}. Analysis of the correlation between various hyperparameters and the validation MAE yields the following architectural insights:
\begin{itemize}
  \item \textbf{Depth Saturation:} Increasing network depth beyond eight layers yielded diminishing returns and occasional instability, with no statistically significant reduction in reconstruction error ($r \approx 0.06$). Consequently, we fixed the encoder depth to the most representative configuration ($L_e=5$) for the exploratory Stage~2 to keep the search space tractable.
  \item \textbf{Activation Robustness:} The tanh activation function with a significantly lower mean validation MAE ($0.0043$) consistently outperformed unbounded alternatives like ReLU ($0.0122$) for the spectral reconstruction task. Its bounded range appears well-suited for modelling the normalised spectral intensities, making it a primary candidate.
  \item \textbf{Input Representation:} The real datatype (mean validation MAE of $0.0033$) significantly outperformed imaginary ($0.0042$) and magnitude ($0.0085$) inputs across all configurations, justifying its exclusive use in the final model.
\end{itemize}

\textbf{Stage~2: Quantifier Capacity Estimation (Frozen Encoder)} With the encoder architecture fixed to a high-performing configuration from Stage~1, we conducted a targeted search to determine the necessary capacity for the quantifier module.

Sensitivity analysis of the quantifier branch shows the risks of over-parameterisation of the search space, detailed in Table~\ref{tab:quan-para}. Full selection results can be found in the supplementary materials~\cite{repo-model-ae}.
\begin{itemize}
  \item \textbf{Width Constraint (Smaller is Better):} Analysis of the median validation MAE reveals a clear positive trend between layer width and error. Models with compact hidden layers ($128$--$256$ units) consistently achieved lower median errors (MAE $\approx 0.024$) compared to wider networks (e.g., $2048$ units, MAE $\approx 0.030$).
  \item \textbf{Activation Function Dominance:} We see a strong preference for the sigmoid activation function in the quantifier module. For the hidden layers, models utilising sigmoid achieved the lowest minimum validation MAE ($0.0171$) compared to ReLU ($0.0200$) and tanh ($0.0207$), indicating a higher performance ceiling for this specific regression task. For the output layer, the advantage of bounded activations was even more pronounced. The sigmoid output activation achieved a median MAE of $0.0246$, outperforming unbounded linear ($0.0280$) and ReLU ($0.2615$) activation. Overall, this supports the use of bounded activation functions to match the normalised range of metabolite concentrations.
\end{itemize}

\begin{table}
  \caption{Model selection parameters for quantifier path}\label{tab:quan-para}
  \centering
  \begin{tabular}{lll}
    \textbf{Group} & \textbf{Parameter} & \textbf{Values}                                     \\ \hline
    Dataset        & Normalisation      & sum                                                 \\
    & Acquisitions       & (DIFF, OFF, ON), (DIFF, ON), (DIFF, OFF), (OFF, ON) \\
    & Datatypes          & (real)                                              \\ \hline
    Model          & $N_q$              & $2048, 1024, 512, 384, 256, 192, 128$               \\
    & $L_q$              & $2, 3, 4, 5$                                        \\
    & $a_q$              & linear, ReLU, sigmoid, tanh                         \\
    & $a_m$              & linear, ReLU, sigmoid, softmax, tanh                \\ \hline
    & Fixed Parameters   & $N_f=2048$, $N_m=5$                                 \\ \hline
    Training       & batch size         & $16, 32$                                            \\ \hline
  \end{tabular}
\end{table}

\textbf{Conclusion of Exploration:}

Stages~1 and~2 narrowed the search space (encoder depth, activation choices, quantifier width, etc.). The final joint optimisation (Stage~3, Section~\ref{sec:joint}) was then run over this reduced set of options (Table~\ref{tab:refined_space}).

\begin{table}[!t]
  \caption{\textbf{Refined Search Space} for the final joint optimisation (Stage~3), derived from the insights of Stages 1 \& 2.}\label{tab:refined_space}
  \centering
  \begin{tabular}{lll}
    \textbf{Group} & \textbf{Parameter}            & \textbf{Refined Values (Stage~3)}      \\ \hline
    Dataset        & Normalisation                 & sum                                    \\
    & Acquisitions                  & (DIFF, OFF, ON), (DIFF, ON), (OFF, ON) \\
    & Datatypes                     & (real), (imaginary, real)              \\ \hline
    Model          & Encoder Depth ($L_e$)         & $5, 6$                                 \\
    & Decoder Depth ($L_d$)         & $6, 7, 8$                              \\
    & Encoder Activation ($a_e$)    & tanh, ReLU                             \\
    & Encoder Dropout Rate ($d_e$)  & $0.2, 0.3$                             \\
    & Decoder Activation ($a_d$)    & tanh, linear                           \\
    & Quantifier Width ($N_q$)      & $128, 192, 256, 384, 448, 512$         \\
    & Quantifier Activation ($a_q$) & ReLU, sigmoid                          \\
    & Output Activation ($a_m$)     & sigmoid, softmax                       \\ \hline
  \end{tabular}
\end{table}

\subsubsection{Final Optimisation}\label{sec:joint}

In the final stage, we performed a Bayesian optimisation on the fully, jointly-trained YAE architecture, confined to the refined search space in Table~\ref{tab:refined_space} as identified by Stages~1 and~2 (Section~\ref{sec:archs_yae}). To ensure stable convergence during this end-to-end training, we implemented a dynamic loss-weighting strategy. The total loss $\mathcal{L}_{total}$ is defined as a weighted sum of the reconstruction loss ($\mathcal{L}_{ae}$) and the quantification loss ($\mathcal{L}_{q}$):
\begin{equation}
  \mathcal{L}_{total} = w_{ae} \cdot \mathcal{L}_{ae} + w_{q}(t) \cdot \mathcal{L}_{q}
\end{equation}
where $w_{ae}$ is fixed at $1.0$, but the quantifier weight $w_{q}(t)$ is dynamic. We employed a ``warm-up'' ramp strategy where $w_{q}$ starts at a low value ($0.1$) and linearly increases to its target value ($1.0$) over the initial epochs. Starting with a low $w_{q}$ allows the encoder to learn spectral features from the denoising task first, so that noisy gradients from the untrained quantifier do not disrupt early feature learning. All results of this joint optimisation are provided in~\cite{repo-model-ae}.

\begin{figure}[!t]
  \centering
  \includegraphics[width=.9\textwidth]{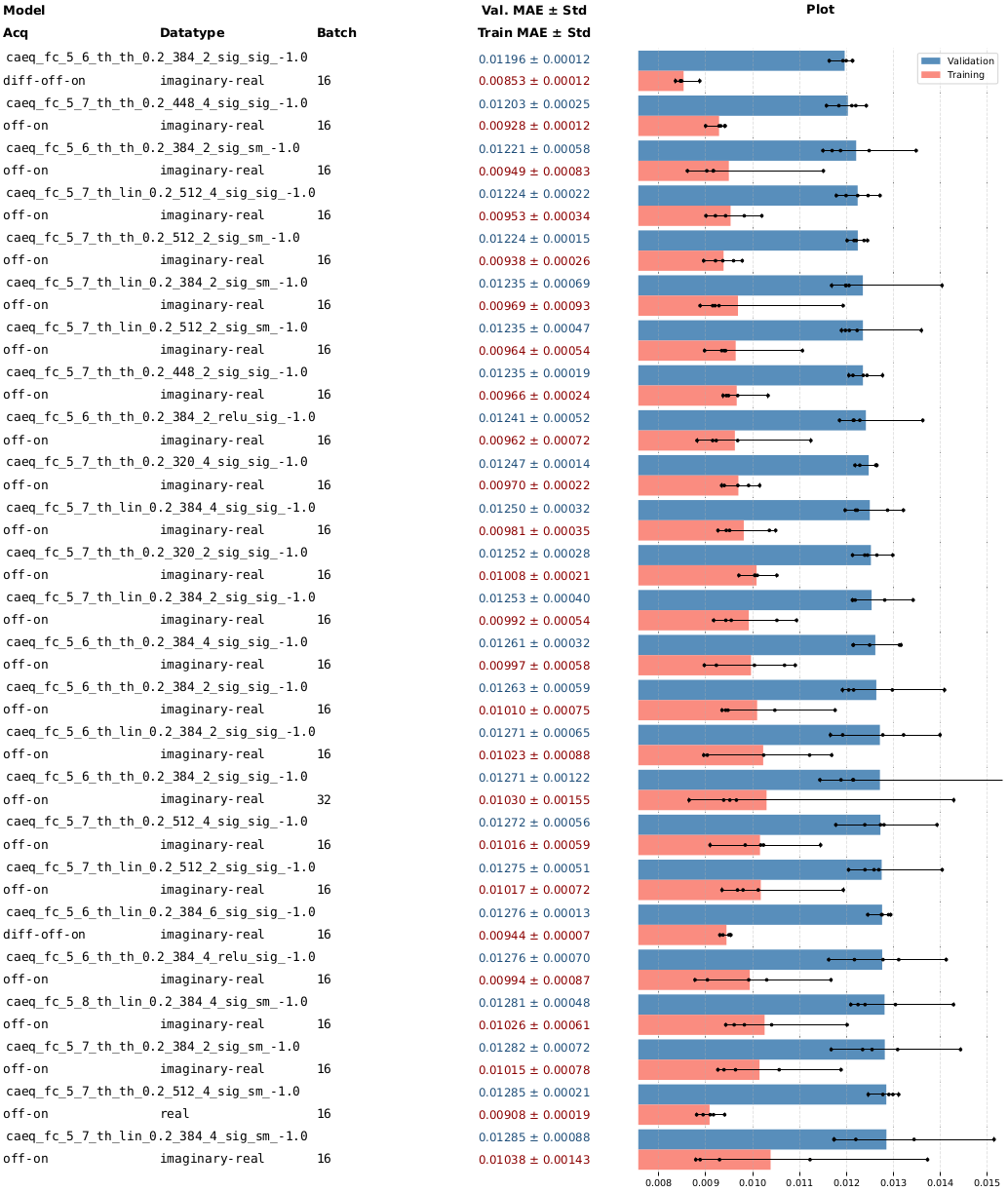}
  \caption{Validation and training concentration MAE for YAE configurations from the final joint optimisation (Stage~3). Each pair of horizontal bars corresponds to one configuration: blue, validation MAE\@; red, training MAE\@. Error bars show standard deviation across five-fold cross-validation. Configurations are ordered by validation MAE (best at top). Dataset: $10{,}000$ simulated spectra, sum normalisation, $200$ epochs (Table~\ref{tab:refined_space}).}\label{fig:caeq_rank}
\end{figure}

\textbf{1. Architectural Convergence and Consistency:}
The joint optimisation results strongly support the architectural insights from the exploratory stages. As summarised in the performance distribution analysis in Figure~\ref{fig:caeq_rank}:
\begin{itemize}
  \item \textbf{Encoder:} The optimal encoder configuration converged to a depth of $L_e=5$ and $L_d=6$, using tanh activations and a dropout rate of $0.2$, confirming the ``middle-ground'' depth hypothesis from Stage~1.
  \item \textbf{Quantifier:} Consistent with Stage~2 findings, the quantifier preferred a compact structure. The top-performing models consistently utilised a width of $384$ units and a depth of two layers.
  \item \textbf{Activation function:} The distinct preference for different activation functions was maintained in the joint setting: tanh for the spectral reconstruction path (encoder/decoder) and sigmoid for the quantification path (hidden and output layers). The results support decoupling the architectural choices for the two paths.
\end{itemize}

\textbf{2. Task-Specific Data Preference:}
A notable finding in this final stage was a shift in the optimal input representation. While Stage~1 (pure denoising) favoured the real datatype, the joint optimisation for quantification accuracy favoured imaginary--real input, particularly with the OFF--ON acquisition pair. This suggests that the real datatype produces better reconstruction results, but the imaginary component carries phase information that helps quantification.

\textbf{3. The Final Model:}
Based on the minimum validation MAE and cross-validation stability across five folds (Figure~\ref{fig:caeq_rank}), the final selected YAE configuration is in Table~\ref{tab:final_yae_config}.

\begin{table}[!t]
  \caption{The final, optimal YAE configuration selected from the joint optimisation Stage~3. This configuration achieved the lowest validation concentration error and demonstrated high stability across five-fold cross-validation.}\label{tab:final_yae_config}
  \centering
  \begin{tabular}{lll}
    \hline
    \textbf{Group} & \textbf{Parameter}        & \textbf{Selected Value}               \\ \hline
    Dataset        & Acquisitions              & \texttt{edit\_off}, \texttt{edit\_on} \\
    & Datatypes                 & \texttt{imaginary}, \texttt{real}     \\
    & Normalisation             & sum                                   \\ \hline
    Encoder        & Depth ($L_e$)             & 5                                     \\
    & Activation ($a_e$)        & tanh                                  \\
    & Dropout Rate ($d_e$)      & 0.2                                   \\ \hline
    Decoder        & Depth ($L_d$)             & 6                                     \\
    & Hidden Activation         & tanh                                  \\
    & Output Activation ($a_d$) & tanh                                  \\ \hline
    Quantifier     & Depth ($L_q$)             & 2                                     \\
    & Width ($N_q$)             & 384                                   \\
    & Hidden Activation ($a_q$) & sigmoid                               \\
    & Output Activation ($a_m$) & sigmoid                               \\
    & Dropout Rate ($d_q$)      & None ($-1.0$)                         \\ \hline
    Training       & Batch Size                & 16                                    \\ \hline
  \end{tabular}
\end{table}

This specific configuration achieved the lowest concentration MAE with low variance shown in Figure~\ref{fig:caeq_rank}, giving the best trade-off between model complexity and generalisation. This is the ``Optimised YAE'' used for the comparative evaluation in Section~\ref{sec:performance_simul} and the experimental validation in Section~\ref{sec:experiments}.

\subsection{Performance of Optimal Configurations on Simulated Data}\label{sec:performance_simul}

Following the optimisation process (Sections~\ref{sec:selection_cnn} and~\ref{sec:selection_yae}), we validated our final, optimised CNN and YAE models against the adapted baseline architectures (FCNN, QNet, QNetBasis, QMRS, and EncDec) using the simulated dataset. The baseline models in Section~\ref{sec:additional_baselines} were implemented using their published parameters rather than performing an equivalent Bayesian optimisation. This methodological choice was twofold: (1) to assess the ``out-of-the-box'' generalisation of these established architectures, and (2) the computational infeasibility of exhaustively optimising all additional models (see also Section~\ref{sec:limitations}).

\begin{figure}[!t]
  \centering
  \includegraphics[width=0.8\textwidth]{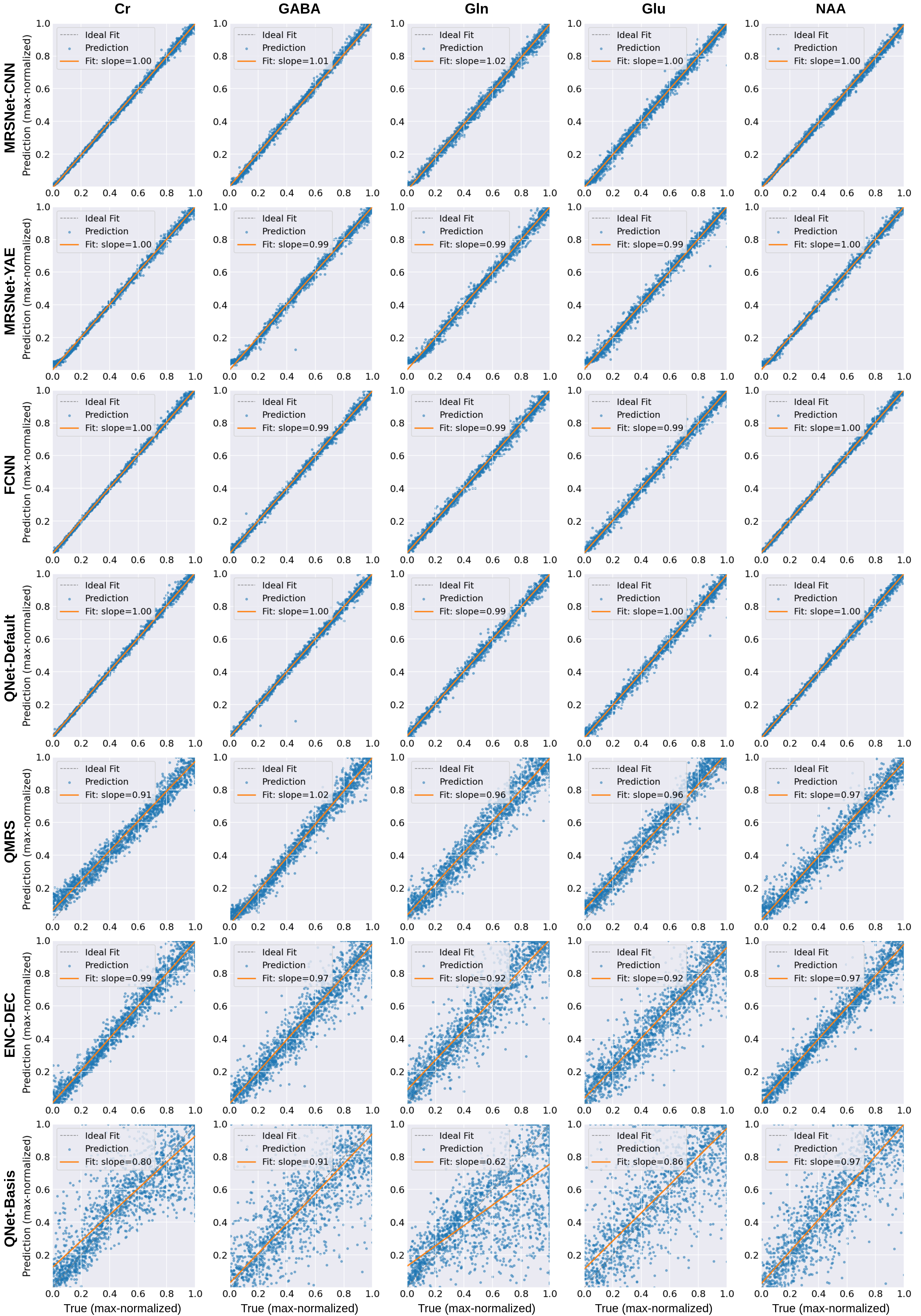}
  \caption{Performance of optimal-configuration models in estimating metabolite concentrations for $2{,}000$ simulated validation spectra. Scatter plots of predicted vs actual (ground-truth) concentrations for each metabolite, with identity line; each point is one spectrum; each panel corresponds to one model and one metabolite (Cr, GABA, Gln, Glu).}\label{fig:simul-comparison}
\end{figure}

Figure~\ref{fig:simul-comparison} and Table~\ref{tab:comparison} summarise the quantification performance of the evaluated models (Training/Validation Split: $80\%$/$20\%$) for ``idealised'' simulated data. Both report the four non-reference metabolites (Cr, GABA, Gln, Glu); NAA is omitted as it is fixed to $1$ by the choice of reference (Section~\ref{sec:preproc}). The scatter plots of the estimated vs actual concentrations of metabolites quantified (Figure~\ref{fig:simul-comparison}) show that for MRSNet-CNN, MRSNet-YAE, FCNN and QNet-default, the points lie close to a line through the origin with slope $1.00$, in most cases to three significant digits. For QMRS, EncDec and QNetBasis, there is significantly more scatter, and the slopes of the linear regression fits deviate more noticeably from $1.00$. Table~\ref{tab:comparison} further shows that for both our systematically optimised models and the simple direct-regression models (FCNN, QNet-default), the $R^2$ values are $\ge 0.99$ and MAEs are small (e.g., $0.024$--$0.026$ for GABA across the four best-performing models; see Table~\ref{tab:comparison}). The more complex baseline models, on the other hand, struggled significantly with the data. QMRS and EncDec exhibited much wider prediction scatter, lower $R^2$ values (e.g., $0.91$ for Gln), and substantially higher MAEs (e.g., Glu MAE of $0.068$ and $0.067$, respectively). The QNetBasis model performed the worst, with the largest errors across metabolites (e.g., Cr $R^2=0.76$, MAE$=0.11$; Gln $R^2=0.82$, MAE$=0.10$).

The results demonstrate that for simple direct-regression models (FCNN, QNet-default), no further optimisation was necessary to achieve near-perfect simulated performance. Conversely, the poor performance of the complex hybrid models (QMRS, EncDec, QNetBasis) suggests they are either less suited to this task or require significant, dataset-specific tuning, indicating their limited ``out-of-the-box'' utility in our specific MEGA-PRESS setup.

\begin{table}[!t]
  \centering
  \caption{Performance indices for optimal-configuration models in estimating metabolite concentrations for $2{,}000$ simulated validation spectra}\label{tab:comparison}
  \begin{subtable}[t]{0.475\textwidth}
    \centering
    \caption{Cr}
    \resizebox{\textwidth}{!}{
      \begin{tabular}{lccc}
        \toprule
        Model        & $R^2$   & SE(slope) & MAE$\pm$Std       \\
        \midrule
        EncDec       & $0.977$ & $0.0034$  & $0.033 \pm 0.037$ \\
        FCNN         & $0.999$ & $0.0008$  & $0.008 \pm 0.009$ \\
        MRSNet-CNN   & $0.999$ & $0.0008$  & $0.009 \pm 0.009$ \\
        MRSNet-YAE   & $0.998$ & $0.0009$  & $0.008 \pm 0.010$ \\
        QMRS         & $0.980$ & $0.0029$  & $0.039 \pm 0.036$ \\
        QNetBasis    & $0.761$ & $0.0101$  & $0.111 \pm 0.116$ \\
        QNet-default & $0.999$ & $0.0009$  & $0.009 \pm 0.011$ \\
        \bottomrule
    \end{tabular}}
  \end{subtable}
  \hspace{0.5cm}
  \begin{subtable}[t]{0.475\textwidth}
    \centering
    \caption{GABA}
    \resizebox{\textwidth}{!}{%
      \begin{tabular}{lccc}
        \toprule
        Model        & $R^2$   & SE(slope) & MAE$\pm$Std       \\
        \midrule
        EncDec       & $0.947$ & $0.0052$  & $0.050 \pm 0.056$ \\
        FCNN         & $0.989$ & $0.0023$  & $0.024 \pm 0.025$ \\
        MRSNet-CNN   & $0.990$ & $0.0022$  & $0.024 \pm 0.025$ \\
        MRSNet-YAE   & $0.985$ & $0.0027$  & $0.026 \pm 0.029$ \\
        QMRS         & $0.960$ & $0.0041$  & $0.040 \pm 0.045$ \\
        QNetBasis    & $0.874$ & $0.0071$  & $0.070 \pm 0.073$ \\
        QNet-default & $0.990$ & $0.0023$  & $0.024 \pm 0.026$ \\
        \bottomrule
    \end{tabular}}
  \end{subtable}
  \vspace{0.7em}
  \begin{subtable}[t]{0.475\textwidth}
    \centering
    \caption{Gln}
    \resizebox{\textwidth}{!}{%
      \begin{tabular}{lccc}
        \toprule
        Model        & $R^2$   & SE(slope) & MAE$\pm$Std       \\
        \midrule
        EncDec       & $0.909$ & $0.0065$  & $0.075 \pm 0.077$ \\
        FCNN         & $0.976$ & $0.0034$  & $0.040 \pm 0.039$ \\
        MRSNet-CNN   & $0.978$ & $0.0032$  & $0.039 \pm 0.038$ \\
        MRSNet-YAE   & $0.970$ & $0.0040$  & $0.043 \pm 0.043$ \\
        QMRS         & $0.925$ & $0.0058$  & $0.069 \pm 0.070$ \\
        QNetBasis    & $0.821$ & $0.0088$  & $0.100 \pm 0.101$ \\
        QNet-default & $0.977$ & $0.0032$  & $0.040 \pm 0.040$ \\
        \bottomrule
    \end{tabular}}
  \end{subtable}
  \hspace{0.5cm}
  \begin{subtable}[t]{0.475\textwidth}
    \centering
    \caption{Glu}
    \resizebox{\textwidth}{!}{
      \begin{tabular}{lccc}
        \toprule
        Model        & $R^2$   & SE(slope) & MAE$\pm$Std       \\
        \midrule
        EncDec       & $0.907$ & $0.0066$  & $0.068 \pm 0.074$ \\
        FCNN         & $0.976$ & $0.0034$  & $0.038 \pm 0.039$ \\
        MRSNet-CNN   & $0.978$ & $0.0033$  & $0.038 \pm 0.039$ \\
        MRSNet-YAE   & $0.971$ & $0.0039$  & $0.040 \pm 0.043$ \\
        QMRS         & $0.912$ & $0.0063$  & $0.067 \pm 0.071$ \\
        QNetBasis    & $0.789$ & $0.0096$  & $0.093 \pm 0.101$ \\
        QNet-default & $0.979$ & $0.0032$  & $0.037 \pm 0.039$ \\
        \bottomrule
    \end{tabular}}
  \end{subtable}
\end{table}

On simulated data, the best models (MRSNet-YAE, MRSNet-CNN, FCNN, QNet-default) are nearly indistinguishable; discrimination requires evaluation on experimental phantoms (Section~\ref{sec:experiments}). The near-perfect performance of the top-four regression models (MRSNet-YAE, MRSNet-CNN, FCNN, QNet-default) establishes a clear performance baseline, but this simulated benchmark lacks the discriminative power to differentiate their performance. The critical test is their performance on experimental phantom data and the sim-to-real challenge (Section~\ref{sec:experiments}). Note that QMRS, EncDec and QNetBasis were originally proposed for different acquisition protocols and more complex \emph{in vivo} settings, and were adapted to our MEGA-PRESS pipeline without extensive re-tuning, so our evaluation reflects their out-of-the-box behaviour in this specific context rather than a reproduction of their originally reported results.

\section{Results and Discussion of Best-Performing Models}\label{sec:experiments}

We evaluate the best-performing DL models (selected on simulated data in Section~\ref{sec:selection}) on the $144$ spectra from $112$ experimental phantoms with known ground-truth concentrations, and compare them to LCModel. LCModel was run on OFF and DIFF spectra separately; at the experiment level, LCModel-OFF achieved significantly lower errors than LCModel-DIFF for GABA and Glu (Table~\ref{tab:lcmodel_off_vs_diff}), and is therefore used as the conventional baseline. We first present the overall phantom results and statistical comparisons, then discuss the sim-to-real gap and its implications.

\subsection{Experimental Ground-Truth Validation}

\begin{figure}[t]
  \centering
  \includegraphics[width=1.0\textwidth]{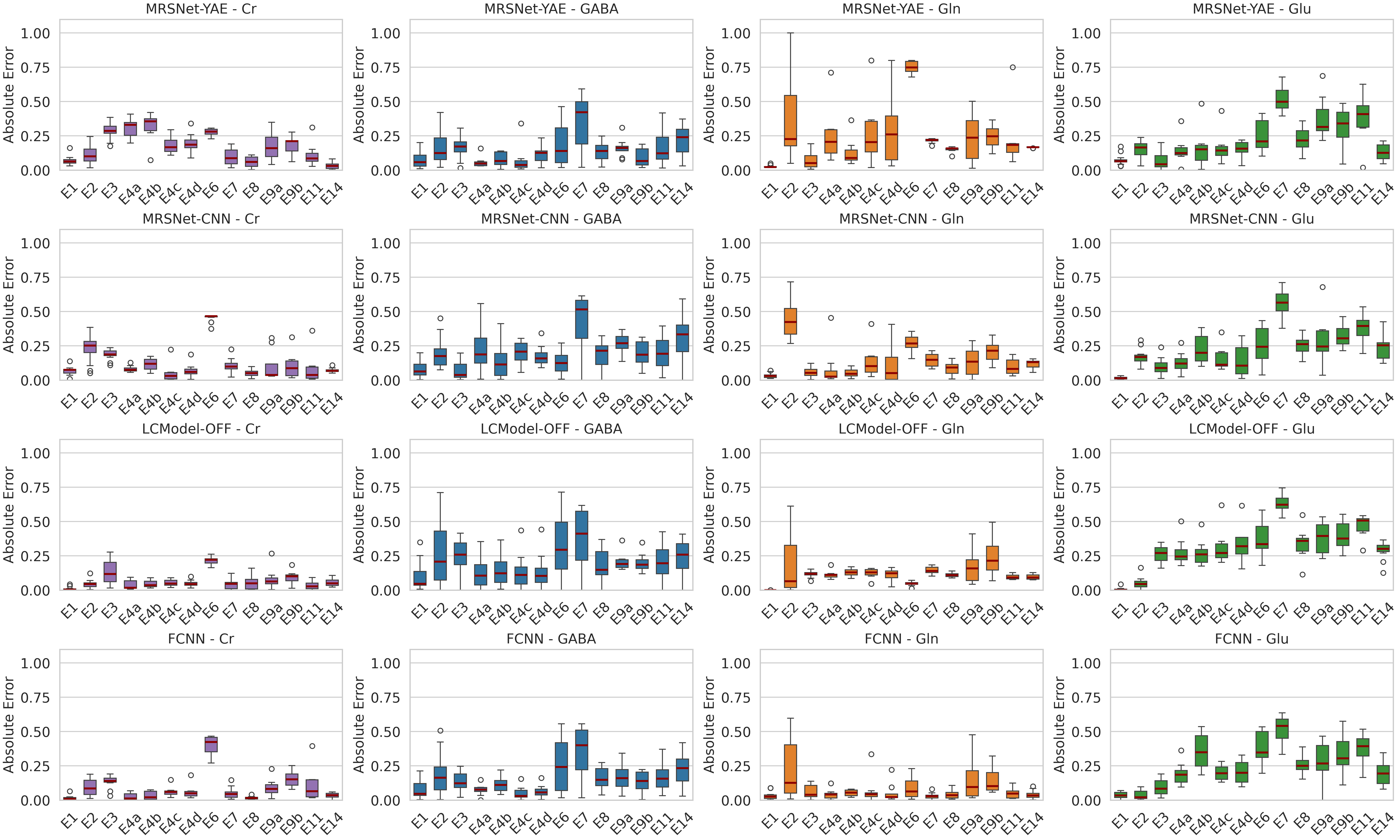}
  \caption{Distribution of absolute quantification errors (max-normalised concentrations) for MRSNet-YAE, MRSNet-CNN, FCNN, and LCModel-OFF across $144$ spectra from $112$ phantoms, grouped by experimental series (E1--E14) and metabolite (Cr, GABA, Gln, Glu). Each box summarises the errors for one series--metabolite--model combination; outliers (points beyond $1.5$ IQR) are shown. NAA is the reference metabolite and is not quantified separately.}\label{fig:mae-box}
\end{figure}

\begin{table}[t]
  \centering
  \caption{Experiment-level comparison between LCModel-OFF and LCModel-DIFF using absolute errors on phantom data. Cr is omitted as it can be quantified from OFF\@; the comparison focuses on GABA, Gln and Glu, for which OFF vs DIFF quantification differs. For each experiment and metabolite, absolute errors were first computed for individual spectra, and the mean error across spectra was used as a single experiment-level summary statistic to avoid pseudo-replication (i.e., all values in this table are based on mean MAE). Paired one-tailed Wilcoxon signed-rank tests were applied across experiments to assess whether LCModel-OFF yields systematically lower errors (Section~\ref{sec:stat_compare}). LCModel-OFF achieved significantly lower errors for GABA ($p=0.045$, lower error in $11$ out of $14$ experiments) and Glu ($p=0.0019$, lower error in $12$ out of $14$ experiments), while no significant difference was observed for Gln. These results indicate that LCModel-OFF provides equal or superior quantification performance compared to LCModel-DIFF at the experiment level, supporting its use as a conservative and stable baseline for subsequent model comparisons.}\label{tab:lcmodel_off_vs_diff}
  \begin{tabular}{lcccc}\hline
    Metabolite                 &
    $p$-value (OFF $<$ DIFF)   &
    Lower error in experiments &
    Mean MAE difference        &
    $N_{\text{exp}}$                                                                \\\hline
    GABA                       & $0.045$  & $11/14$ & $-1.35 \times 10^{-2}$ & $14$ \\
    Gln                        & $0.809$  & $4/14$  & $+3.62 \times 10^{-3}$ & $14$ \\
    Glu                        & $0.0019$ & $12/14$ & $-7.49 \times 10^{-2}$ & $14$ \\\hline
  \end{tabular}
\end{table}

\begin{table}[!t]
  \caption{Phantom performance for optimal-configuration models compared with the LCModel baseline. Experimental series E1--E14 and phantom compositions are summarised in Table~\ref{BenchmarkExperiment}. Values are experiment-level mean MAE $\pm$ SEM (standard error of the mean; max-normalised concentrations).}\label{tab:comparison-experiment}
  \centering
  \small
  \resizebox{\textwidth}{!}{%
    \begin{tabular}{lcccc|cccc}
      \toprule
      Series & \multicolumn{4}{c}{Cr}    & \multicolumn{4}{c}{GABA}                                                                                                                                                                 \\
      & MRSNet-YAE                & MRSNet-CNN                & LCModel-OFF               & FCNN                      & MRSNet-YAE                & MRSNet-CNN                & LCModel-OFF      & FCNN                      \\
      \midrule
      E1     & $0.069 \pm .009$          & $0.067 \pm .009$          & $\mathbf{0.009} \pm .004$ & $0.015 \pm .004$          & $\mathbf{0.079} \pm .017$ & $0.086 \pm .019$          & $0.096 \pm .029$ & $0.081 \pm .020$          \\
      E2     & $0.108 \pm .015$          & $0.229 \pm .023$          & $\mathbf{0.043} \pm .007$ & $0.092 \pm .014$          & $\mathbf{0.171} \pm .029$ & $0.188 \pm .026$          & $0.261 \pm .056$ & $0.185 \pm .036$          \\
      E3     & $0.291 \pm .015$          & $0.188 \pm .009$          & $\mathbf{0.133} \pm .022$ & $0.137 \pm .011$          & $0.167 \pm .020$          & $\mathbf{0.070} \pm .016$ & $0.247 \pm .031$ & $0.132 \pm .019$          \\
      E4a    & $0.305 \pm .027$          & $0.081 \pm .009$          & $0.038 \pm .012$          & $\mathbf{0.026} \pm .009$ & $\mathbf{0.059} \pm .015$ & $0.236 \pm .065$          & $0.133 \pm .046$ & $0.074 \pm .015$          \\
      E4b    & $0.317 \pm .039$          & $0.115 \pm .016$          & $0.045 \pm .009$          & $\mathbf{0.032} \pm .011$ & $\mathbf{0.076} \pm .019$ & $0.141 \pm .049$          & $0.145 \pm .045$ & $0.115 \pm .022$          \\
      E4c    & $0.185 \pm .025$          & $\mathbf{0.052} \pm .026$ & $0.053 \pm .009$          & $0.063 \pm .014$          & $0.075 \pm .039$          & $0.199 \pm .031$          & $0.139 \pm .050$ & $\mathbf{0.052} \pm .018$ \\
      E4d    & $0.194 \pm .028$          & $0.071 \pm .019$          & $\mathbf{0.050} \pm .009$ & $0.060 \pm .018$          & $0.113 \pm .024$          & $0.179 \pm .029$          & $0.138 \pm .051$ & $\mathbf{0.064} \pm .018$ \\
      E6     & $0.276 \pm .009$          & $0.452 \pm .010$          & $\mathbf{0.213} \pm .009$ & $0.398 \pm .022$          & $0.188 \pm .052$          & $\mathbf{0.128} \pm .026$ & $0.326 \pm .076$ & $0.256 \pm .062$          \\
      E7     & $0.100 \pm .015$          & $0.103 \pm .012$          & $\mathbf{0.040} \pm .008$ & $0.047 \pm .010$          & $\mathbf{0.352} \pm .049$ & $0.424 \pm .050$          & $0.380 \pm .051$ & $0.357 \pm .044$          \\
      E8     & $0.059 \pm .014$          & $0.052 \pm .011$          & $0.056 \pm .019$          & $\mathbf{0.017} \pm .004$ & $\mathbf{0.135} \pm .029$ & $0.183 \pm .039$          & $0.179 \pm .044$ & $0.157 \pm .032$          \\
      E9a    & $0.175 \pm .039$          & $0.102 \pm .042$          & $\mathbf{0.083} \pm .029$ & $0.091 \pm .024$          & $\mathbf{0.166} \pm .025$ & $0.262 \pm .029$          & $0.208 \pm .024$ & $0.166 \pm .036$          \\
      E9b    & $0.179 \pm .027$          & $0.109 \pm .039$          & $\mathbf{0.094} \pm .020$ & $0.155 \pm .023$          & $\mathbf{0.095} \pm .027$ & $0.196 \pm .039$          & $0.202 \pm .029$ & $0.135 \pm .032$          \\
      E11    & $0.111 \pm .032$          & $0.083 \pm .042$          & $\mathbf{0.034} \pm .011$ & $0.111 \pm .045$          & $\mathbf{0.167} \pm .050$ & $0.206 \pm .049$          & $0.208 \pm .050$ & $0.174 \pm .041$          \\
      E14    & $\mathbf{0.032} \pm .007$ & $0.072 \pm .005$          & $0.057 \pm .009$          & $0.035 \pm .005$          & $\mathbf{0.217} \pm .035$ & $0.312 \pm .054$          & $0.239 \pm .039$ & $0.225 \pm .038$          \\
      \bottomrule
  \end{tabular}}%
  \vspace{1.5em}
  \resizebox{\textwidth}{!}{%
    \begin{tabular}{lcccc|cccc}
      \toprule
      Series & \multicolumn{4}{c}{Gln} & \multicolumn{4}{c}{Glu}                                                                                                                                                                           \\
      & MRSNet-YAE              & MRSNet-CNN                & LCModel-OFF               & FCNN                      & MRSNet-YAE                & MRSNet-CNN                & LCModel-OFF               & FCNN                      \\
      \midrule
      E1     & $0.026 \pm .002$        & $0.032 \pm .005$          & $\mathbf{0.000} \pm .000$ & $0.032 \pm .008$          & $0.074 \pm .011$          & $0.015 \pm .002$          & $\mathbf{0.006} \pm .003$ & $0.036 \pm .006$          \\
      E2     & $0.372 \pm .076$        & $0.452 \pm .034$          & $\mathbf{0.166} \pm .048$ & $0.224 \pm .053$          & $0.152 \pm .015$          & $0.166 \pm .014$          & $0.048 \pm .010$          & $\mathbf{0.034} \pm .007$ \\
      E3     & $0.073 \pm .014$        & $\mathbf{0.057} \pm .009$ & $0.116 \pm .005$          & $0.063 \pm .013$          & $\mathbf{0.065} \pm .015$ & $0.095 \pm .015$          & $0.263 \pm .015$          & $0.095 \pm .014$          \\
      E4a    & $0.250 \pm .072$        & $0.090 \pm .054$          & $0.113 \pm .011$          & $\mathbf{0.042} \pm .014$ & $0.146 \pm .035$          & $\mathbf{0.128} \pm .028$ & $0.279 \pm .037$          & $0.193 \pm .030$          \\
      E4b    & $0.130 \pm .037$        & $\mathbf{0.053} \pm .012$ & $0.128 \pm .010$          & $0.053 \pm .009$          & $\mathbf{0.163} \pm .051$ & $0.229 \pm .039$          & $0.273 \pm .035$          & $0.359 \pm .049$          \\
      E4c    & $0.267 \pm .089$        & $0.135 \pm .044$          & $0.123 \pm .013$          & $\mathbf{0.072} \pm .038$ & $0.165 \pm .041$          & $\mathbf{0.157} \pm .032$ & $0.314 \pm .047$          & $0.200 \pm .020$          \\
      E4d    & $0.294 \pm .093$        & $0.117 \pm .054$          & $0.112 \pm .016$          & $\mathbf{0.054} \pm .026$ & $0.150 \pm .022$          & $\mathbf{0.140} \pm .041$ & $0.334 \pm .049$          & $0.208 \pm .033$          \\
      E6     & $0.751 \pm .013$        & $0.269 \pm .019$          & $\mathbf{0.046} \pm .005$ & $0.092 \pm .026$          & $\mathbf{0.247} \pm .036$ & $0.250 \pm .044$          & $0.382 \pm .040$          & $0.384 \pm .038$          \\
      E7     & $0.216 \pm .004$        & $0.146 \pm .011$          & $0.143 \pm .006$          & $\mathbf{0.029} \pm .005$ & $\mathbf{0.510} \pm .021$ & $0.554 \pm .024$          & $0.632 \pm .015$          & $0.520 \pm .024$          \\
      E8     & $0.150 \pm .008$        & $0.085 \pm .018$          & $0.110 \pm .006$          & $\mathbf{0.042} \pm .012$ & $\mathbf{0.227} \pm .033$ & $0.252 \pm .029$          & $0.339 \pm .044$          & $0.260 \pm .027$          \\
      E9a    & $0.237 \pm .062$        & $\mathbf{0.138} \pm .038$ & $0.169 \pm .043$          & $0.154 \pm .058$          & $0.381 \pm .056$          & $0.288 \pm .067$          & $0.381 \pm .041$          & $\mathbf{0.277} \pm .053$ \\
      E9b    & $0.243 \pm .036$        & $0.206 \pm .032$          & $0.244 \pm .055$          & $\mathbf{0.147} \pm .038$ & $\mathbf{0.314} \pm .061$ & $0.323 \pm .033$          & $0.399 \pm .042$          & $0.338 \pm .058$          \\
      E11    & $0.227 \pm .077$        & $0.097 \pm .021$          & $0.097 \pm .007$          & $\mathbf{0.050} \pm .013$ & $0.380 \pm .063$          & $0.378 \pm .040$          & $0.468 \pm .030$          & $\mathbf{0.376} \pm .039$ \\
      E14    & $0.167 \pm .001$        & $0.116 \pm .010$          & $0.097 \pm .006$          & $\mathbf{0.039} \pm .009$ & $\mathbf{0.129} \pm .018$ & $0.239 \pm .030$          & $0.290 \pm .021$          & $0.194 \pm .026$          \\
      \bottomrule
  \end{tabular}}%
\end{table}

\begin{table}[t]
  \centering
  \caption{Overall MAE over all $144$ phantom spectra (max-normalised concentrations). Primary target: GABA\@. Computed from raw per-spectrum absolute errors (DL models from model-dist; LCModel-OFF from LCM-results analysis on the same $144$ spectra). Values are mean $\pm$ SEM with $95\%$ CI (normal approximation) in brackets. The ``Overall'' row is the mean over all $144$ spectra of the per-spectrum MAE (mean over the five metabolites Cr, GABA, Gln, Glu, NAA).}\label{tab:overall_mae}
  \small
  \resizebox{\textwidth}{!}{%
    \begin{tabular}{lcccc}
      \toprule
      Metabolite & MRSNet-YAE                                  & MRSNet-CNN                         & FCNN                                        & LCModel-OFF                                 \\
      \midrule
      Cr         & $0.165 \pm 0.009$ [$.146$, $.183$]          & $0.136 \pm 0.010$ [$.116$, $.155$] & $0.091 \pm 0.009$ [$.074$, $.109$]          & $\mathbf{0.067} \pm 0.006$ [$.056$, $.078$] \\
      GABA       & $\mathbf{0.161} \pm 0.011$ [$.139$, $.183$] & $0.203 \pm 0.013$ [$.178$, $.229$] & $0.167 \pm 0.012$ [$.144$, $.190$]          & $0.220 \pm 0.015$ [$.191$, $.249$]          \\
      Gln        & $0.238 \pm 0.019$ [$.201$, $.276$]          & $0.153 \pm 0.012$ [$.129$, $.177$] & $\mathbf{0.080} \pm 0.009$ [$.062$, $.098$] & $0.116 \pm 0.008$ [$.101$, $.132$]          \\
      Glu        & $\mathbf{0.219} \pm 0.014$ [$.191$, $.246$] & $0.230 \pm 0.014$ [$.201$, $.258$] & $0.237 \pm 0.015$ [$.208$, $.266$]          & $0.304 \pm 0.016$ [$.272$, $.336$]          \\
      \midrule
      Overall    & $0.160 \pm 0.007$ [$.147$, $.173$]          & $0.147 \pm 0.006$ [$.134$, $.159$] & $\mathbf{0.117} \pm 0.006$ [$.105$, $.129$] & $0.144 \pm 0.006$ [$.131$, $.156$]          \\
      \bottomrule
  \end{tabular}}%
\end{table}

Figure~\ref{fig:mae-box} summarises the performance of the three models that performed best on simulated data (MRSNet-YAE, MRSNet-CNN, FCNN) and LCModel on the $144$ spectra ($112$ phantoms). LCModel can only quantify a single spectrum. We therefore quantified OFF and DIFF spectra separately as described in Section~\ref{sec:data} and compared the quantification accuracy. The results of the statistical tests (described in Section~\ref{sec:stat_compare}) are summarised in Table~\ref{tab:lcmodel_off_vs_diff}. The comparison indicates that the LCModel quantification results based on the OFF spectra were generally more accurate than those based on difference spectra, including for GABA\@. This is surprising at first, as quantifying GABA from unedited OFF spectra at $\SI{3}{T}$ is unreliable due to overlap with Cr and macromolecules, and MEGA-PRESS was specifically designed to produce difference spectra to address these issues. However, LCModel was originally designed for normal PRESS or STEAM spectra~\cite{LCModel1993, LCModel2001}; the user manual documents MEGA-PRESS and off-resonance spectra as special cases~\cite[Sections~9.4 and~11.3.1]{LCModelManual}, and there may be issues with the adaptation to the difference spectra. The absence of a macromolecule signal in the phantom data may also play a role here. In any case, as our statistical analysis suggests that LCModel quantification using the OFF spectra is more accurate, we use LCModel-OFF as the conventional baseline in the following comparisons.

Relative to their near-perfect performance on simulated validation data, both the best YAE and best CNN showed substantially higher errors on phantoms. Their accuracy was comparable to, and in some cases worse than, LCModel-OFF, and there is no clear winner. Table~\ref{tab:comparison-experiment} reports experiment-level mean MAE for each of the $14$ series. From the raw per-spectrum absolute errors, the overall MAE (averaged over all $144$ spectra), with standard error of the mean (SEM) and $95\%$ CI, is summarised in Table~\ref{tab:overall_mae}: for GABA (primary target), MAE was $0.161$ (MRSNet-YAE), $0.203$ (MRSNet-CNN), $0.167$ (FCNN), and $0.220$ (LCModel-OFF). By metabolite, LCModel-OFF had the lowest MAE for Cr in $9$ of $14$ series, FCNN for Gln in $9$ series, MRSNet-YAE for GABA in $10$ series and for Glu in $7$ series. Ranking of the three DL methods was unchanged when summarising per-series performance by the median of the $14$ experiment-level MAEs instead of the MAE pooled over all $144$ spectra. Although FCNN achieved the lowest overall MAE on the phantom spectra, cross-validation on simulated data showed consistently higher errors on validation than on training sets (validation MAE $\approx 1.5\times$ training MAE), indicating mild overfitting; the apparently favourable phantom performance should therefore be interpreted with caution rather than as evidence of superior generalisation. Nevertheless, the YAE and CNN models remain competitive on phantom data: MRSNet-YAE achieved the lowest mean MAE for GABA (the primary target) and for Glu, and FCNN for Gln (Table~\ref{tab:overall_mae}). Cr and NAA have well-separated resonances in OFF spectra and do not require edited spectroscopy for quantification. LCModel, designed for standard PRESS or STEAM spectra, can therefore quantify them more simply and accurately from OFF, which is consistent with LCModel-OFF having the lowest MAE for Cr.

Pairwise one-tailed Wilcoxon signed-rank tests (Section~\ref{sec:stat_compare}) were used to test whether one method had systematically lower experiment-level MAE than another. For Cr, the null hypothesis that the other methods were not worse than LCModel-OFF was rejected for MRSNet-YAE and MRSNet-CNN ($\alpha=0.01$) but not for FCNN\@. For GABA and Glu, the null hypothesis that the other methods were not worse than MRSNet-YAE was rejected for LCModel-OFF but not for MRSNet-CNN or FCNN\@. For Gln, the null hypothesis that the other methods were not worse than FCNN was rejected for all other methods. Thus, each of the four methods was statistically superior to at least one other on at least one metabolite, but no method was superior to all others on all metabolites.

Figure~\ref{fig:mae-box} shows that error distributions varied by series, metabolite, and model. Absolute errors for GABA spanned a wider range than for Cr, consistent with Cr's strong, well-separated signal and GABA's overlap with other metabolites. Performance did not follow a simple ordering by phantom type: the deliberately miscalibrated solution series E2 did not consistently show higher errors than the well-calibrated series E3 (e.g., for Cr and GABA, errors in E2 were similar to or lower than in E3 except for MRSNet-CNN); only for Gln did E2 show noticeably larger MAE and spread. Similarly, tissue-mimicking gel phantoms did not consistently yield worse quantification than solution phantoms. Domain shift therefore appears to depend on factors beyond calibration or physical state alone; characterising them is left for future work. Overall, each method had strengths on specific metabolites or conditions, but no single method emerged as best across all scenarios.

\subsection{Impact of Linewidth Augmentation}\label{sec:linewidth_aug}

To test if the sim-to-real gap is primarily driven by the fixed linewidth in our training data, we trained additional versions of the CNN, FCNN and YAE models on a dataset augmented with variable linewidths sampled from a uniform distribution ranging from $1$ to $10$ Hz in $0.2$ Hz steps. This range spans all estimated linewidths in our phantom data (Section~\ref{sec:data_sim}) and was chosen to cover the full range of experimental conditions, from high-quality phantom scans (typically $2$--$4$ Hz) to \emph{in vivo} scenarios where poor shimming can lead to significantly broader lines.

\begin{table}[t]
  \centering
  \caption{MAE over all $144$ phantom spectra for models trained with variable linewidth augmentation ($1$--$10$ Hz). Values are max-normalised relative concentrations (mean $\pm$ SEM with $95\%$ CI in brackets). Bold indicates the lowest error per metabolite. The ``Overall'' row is the mean over all $144$ spectra of the per-spectrum MAE (mean over the five metabolites Cr, GABA, Gln, Glu, NAA).}\label{tab:linewidth_aug_mae}
  \small
  \resizebox{\textwidth}{!}{%
    \begin{tabular}{lcccc}
      \toprule
      Metabolite & MRSNet-YAE (Aug)                            & MRSNet-CNN (Aug)                   & FCNN (Aug)                                  & LCModel-OFF                                 \\
      \midrule
      Cr         & $0.103 \pm 0.008$ [$.088$, $.118$]          & $0.129 \pm 0.009$ [$.111$, $.147$] & $0.102 \pm 0.009$ [$.084$, $.120$]          & $\mathbf{0.067} \pm 0.006$ [$.056$, $.078$] \\
      GABA       & $\mathbf{0.151} \pm 0.010$ [$.132$, $.170$] & $0.161 \pm 0.011$ [$.140$, $.182$] & $0.160 \pm 0.011$ [$.139$, $.181$]          & $0.220 \pm 0.015$ [$.191$, $.249$]          \\
      Gln        & $0.194 \pm 0.020$ [$.155$, $.234$]          & $0.151 \pm 0.013$ [$.126$, $.176$] & $\mathbf{0.099} \pm 0.011$ [$.079$, $.120$] & $0.116 \pm 0.008$ [$.101$, $.132$]          \\
      Glu        & $\mathbf{0.184} \pm 0.015$ [$.156$, $.213$] & $0.212 \pm 0.014$ [$.185$, $.239$] & $0.209 \pm 0.014$ [$.183$, $.236$]          & $0.304 \pm 0.016$ [$.272$, $.336$]          \\
      \midrule
      Overall    & $0.131 \pm 0.008$ [$.115$, $.146$]          & $0.133 \pm 0.005$ [$.122$, $.143$] & $\mathbf{0.116} \pm 0.005$ [$.106$, $.127$] & $0.144 \pm 0.006$ [$.131$, $.156$]          \\
      \bottomrule
  \end{tabular}}%
\end{table}

Table~\ref{tab:linewidth_aug_mae} summarises the linewidth-augmented results with the non-augmented results in Table~\ref{tab:overall_mae}. All augmented models maintained near-perfect performance on the simulated test set (per-metabolite MAEs on the order of $10^{-2}$). On the experimental phantoms, augmentation led to performance improvements across architectures. For GABA, the augmented YAE achieved the lowest error ($0.151$), followed by FCNN ($0.160$) and CNN ($0.161$), all outperforming LCModel-OFF ($0.220$). For Glu, the augmented YAE also achieved the lowest error ($0.184$), followed by FCNN ($0.209$) and CNN ($0.212$), all improving upon their non-augmented baselines and LCModel ($0.304$). For Gln, the FCNN remained the best performing model ($0.099$), although this was worse than its non-augmented performance ($0.080$), while the augmented YAE improved ($0.194$) relative to its non-augmented baseline ($0.238$). Note that the best model for each metabolite and overall remain the same in the augmented vs.\ the non-augmented versions.

Despite the gains, the error on phantom data remains an order of magnitude higher than on simulated data ($\approx 0.15$ vs values on the order of $10^{-2}$ on simulations). So while linewidth mismatch is an important factor, it is not the only one. As the gap persists across the three architectures, other unmodelled effects (e.g.\ lineshape asymmetries, baseline instabilities, phase errors) must be addressed in future simulation pipelines, which is well beyond the scope of this study.

\subsection{Discussion}

We performed systematic model selection and ground-truth evaluation of deep learning architectures for MEGA-PRESS quantification. Our results show that DL can match or exceed LCModel on phantoms for GABA and Glu when linewidth augmentation is used, however, a sim-to-real gap remains and performance on simulations alone is a poor predictor of phantom performance.

\subsubsection{Principal Findings}

The principal finding of this work is the stark contrast between model performance on simulated versus experimental data. Following systematic Bayesian optimisation, the final YAE and CNN models achieved near-perfect quantification on a large, independent simulated validation dataset (per-metabolite MAEs on the order of $0.008$ for Cr and $0.024$--$0.026$ for GABA\@; regression slopes and $R^2 \approx 1.00$; see Table~\ref{tab:comparison}), indicating that they had learned the mapping from idealised spectra to metabolite concentrations.

On the $144$ spectra from $112$ experimental phantoms with known ground truth, this high performance was not maintained. Errors increased substantially for both DL models, and neither showed a consistent, statistically significant advantage over LCModel-OFF (Section~\ref{sec:experiments}). Performance was comparable across methods (Table~\ref{tab:comparison-experiment}). No single method was statistically best across all four metabolites. Perhaps most notably, the degradation was evident even on solution-based phantoms, which are closest to the simulation ideal, and not only on tissue-mimicking gels. Cross-validation on simulated data revealed mild but systematic overfitting in the three best-performing deep models (YAE, CNN, FCNN), with validation MAE typically $20$--$70\%$ higher than training MAE across folds, and somewhat stronger overfitting in QNet and QNetBasis, whereas the EncDec model showed little train--validation discrepancy. However, these effects were small compared to the much larger increase in error when moving from simulated data to experimental phantoms. This shows that conventional overfitting on simulations alone cannot account for the sim-to-real degradation.

Variable linewidth augmentation (Section~\ref{sec:linewidth_aug}) reduced the sim-to-real gap: nearly all DL-model performances were improved per metabolite and overall, even if their relative performance remains quite consistent. However, a significant sim-to-real gap remains, and we cannot distinguish how much of this gap is due to other unmodelled factors versus fundamental limitations. Further experiments with more aggressive augmentation and substantially improved simulation pipelines are needed. The relative strength of DL for GABA and Glu/Gln (versus LCModel for Cr) aligns with the fact that edited MEGA-PRESS is most beneficial for overlapping, low-SNR metabolites such as GABA and Glu/Gln, whereas reference metabolites such as Cr are more easily quantified from OFF spectra alone.

Furthermore, our results show that all methods, including LCModel, struggled with out-of-distribution data. The highest quantification errors were observed in experiments E7, E9, and E11, which featured extreme GABA/NAA or Glu/NAA concentration ratios. These conditions represent cases that were likely underrepresented in, or outside the bounds of, the models' training data, emphasising the vulnerability of data-driven models when faced with novel biochemical conditions. While Sobol sampling ensures uniform coverage of the hypercube of relative concentrations, specific physiological or experimental combinations (e.g., extremely high GABA with low NAA) may still fall into sparse regions of the training distribution, contributing to the performance gap.

\subsubsection{Comparison with Literature Baselines}

The comparative deep learning baselines (FCNN, QNet, QMRS, EncDec) were not originally designed for MEGA-PRESS and would require sequence-specific adaptation and optimisation to reach their full potential. They were used in their published configurations with minimal adaptation to our MEGA-PRESS pipeline (Section~\ref{sec:additional_baselines}) and were not subjected to the same Bayesian optimisation as the CNN and YAE\@. Our comparison should therefore be interpreted as assessing the \emph{transferability} of these architectures to MEGA-PRESS quantification, i.e., how well they perform out-of-the-box, rather than their absolute performance limits or optimised potential. They nevertheless serve as a useful reference in particular given the strong performance of the FCNN architecture.

\subsubsection{The Sim-to-Real Gap: Potential Causes}

\begin{figure}[!t]
  \centering
  \subfloat[NAA OFF]{\includegraphics[width=0.195\linewidth]{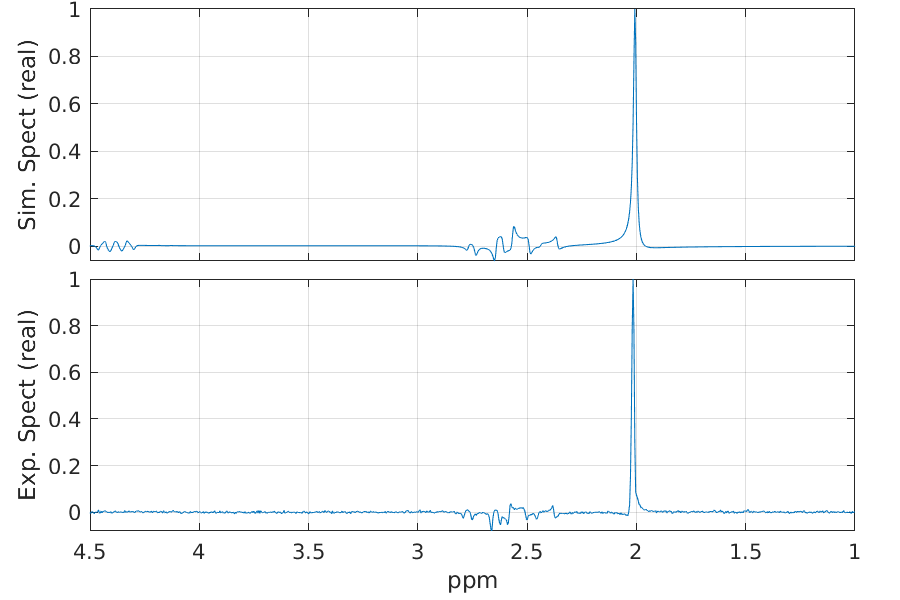}}
  \subfloat[Cr OFF]{\includegraphics[width=0.195\linewidth]{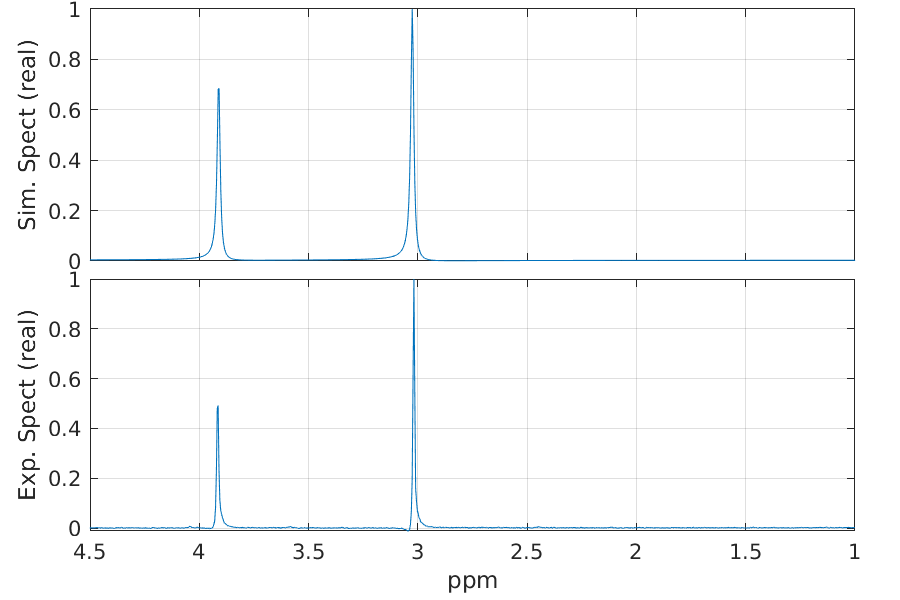}}
  \subfloat[GABA OFF]{\includegraphics[width=0.195\linewidth]{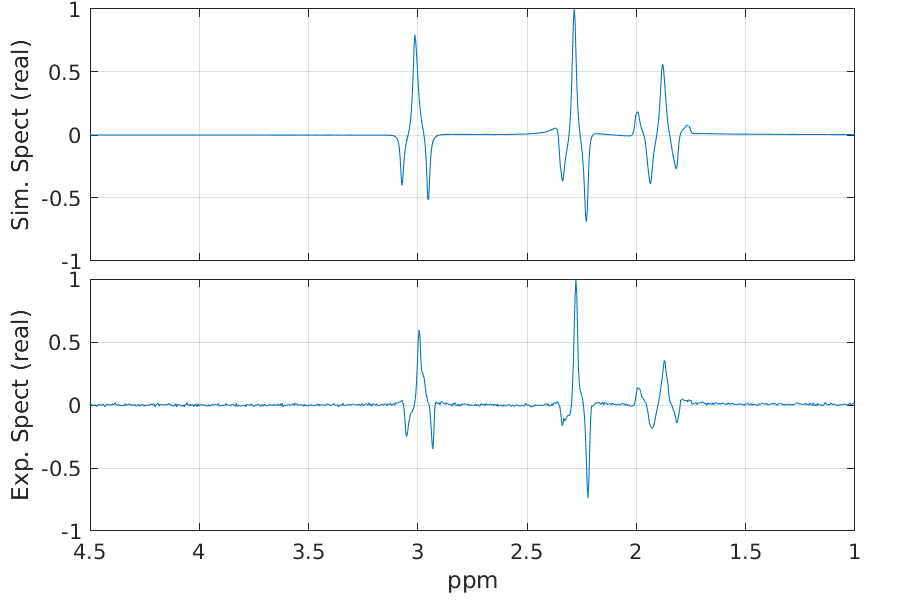}}
  \subfloat[Glu OFF]{\includegraphics[width=0.195\linewidth]{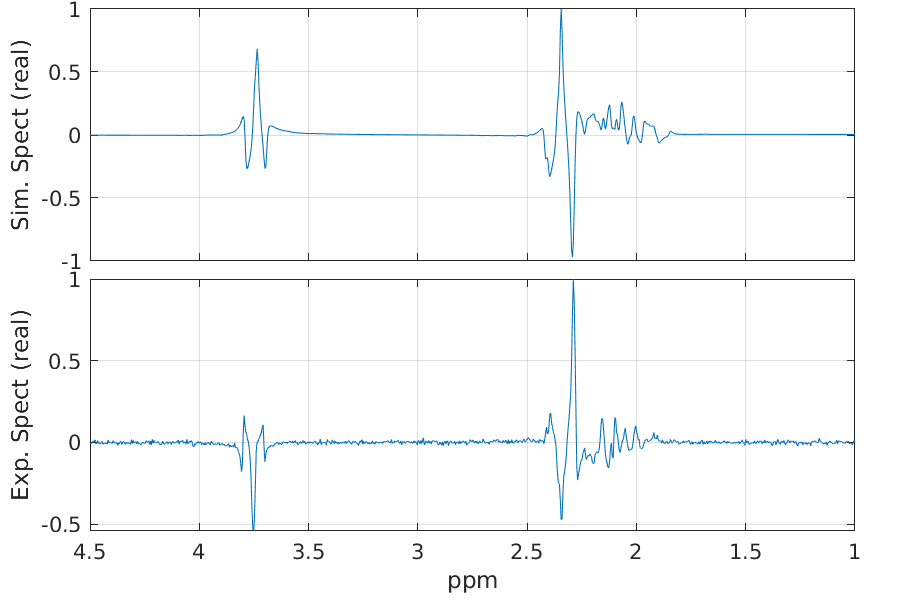}}
  \subfloat[Gln OFF]{\includegraphics[width=0.195\linewidth]{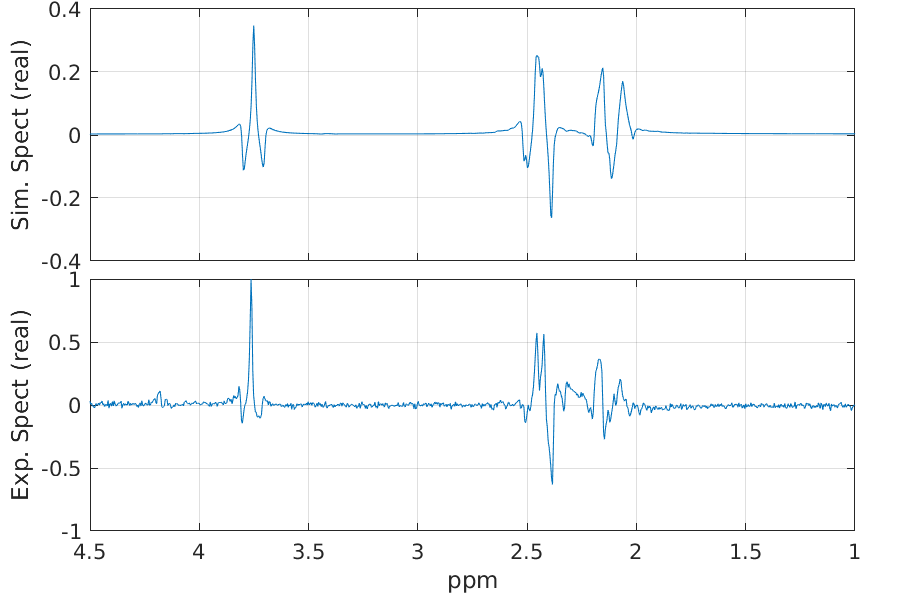}} \\
  \subfloat[NAA ON]{\includegraphics[width=0.195\linewidth]{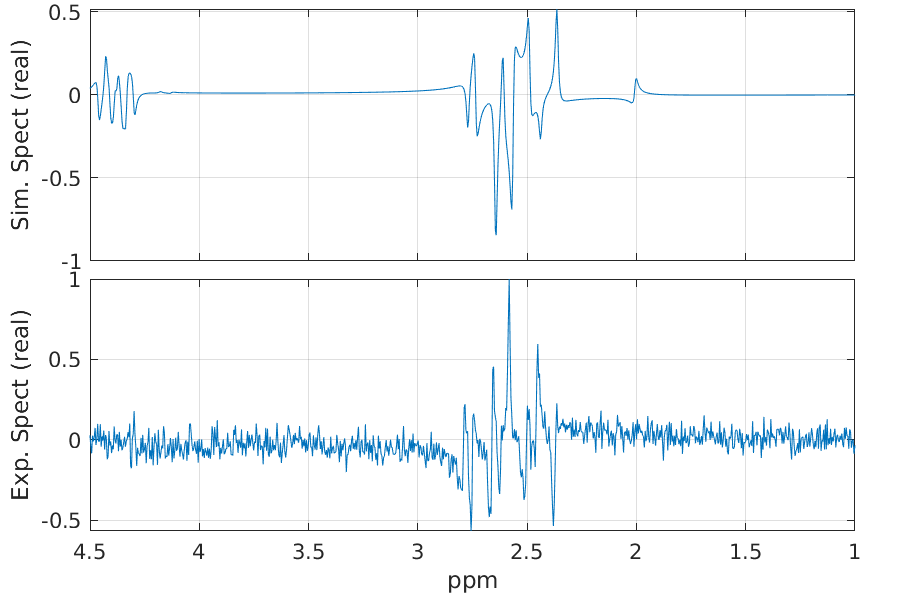}}
  \subfloat[Cr ON]{\includegraphics[width=0.195\linewidth]{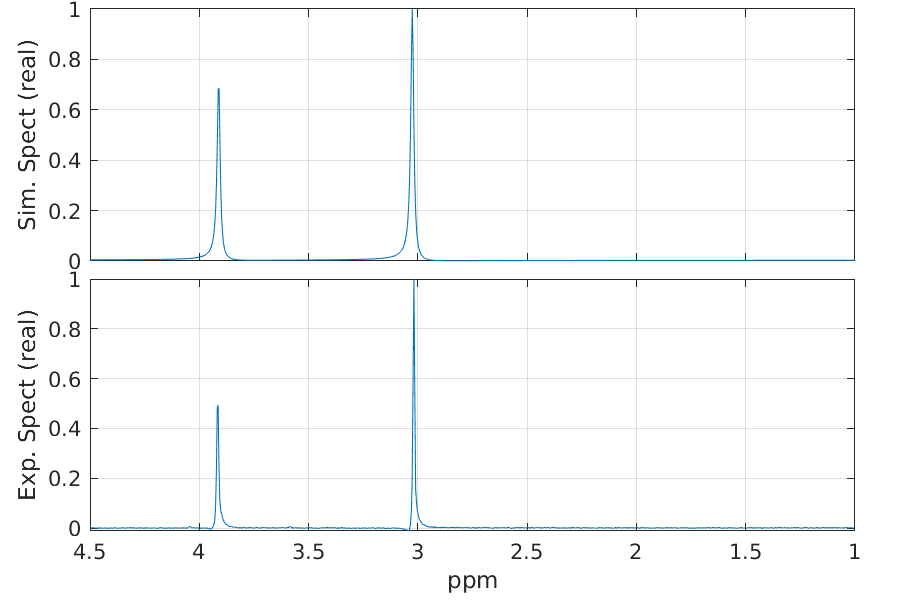}}
  \subfloat[GABA ON]{\includegraphics[width=0.195\linewidth]{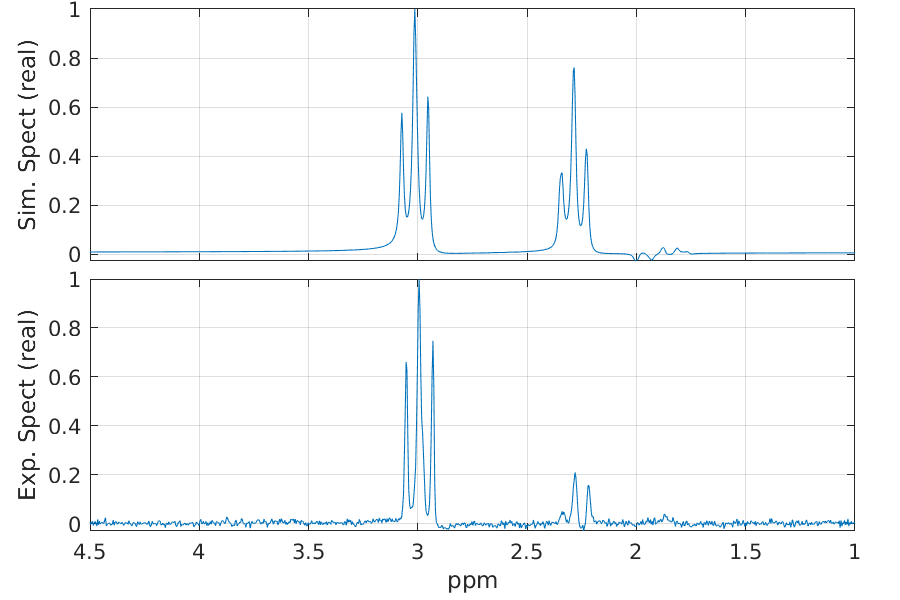}}
  \subfloat[Glu ON]{\includegraphics[width=0.195\linewidth]{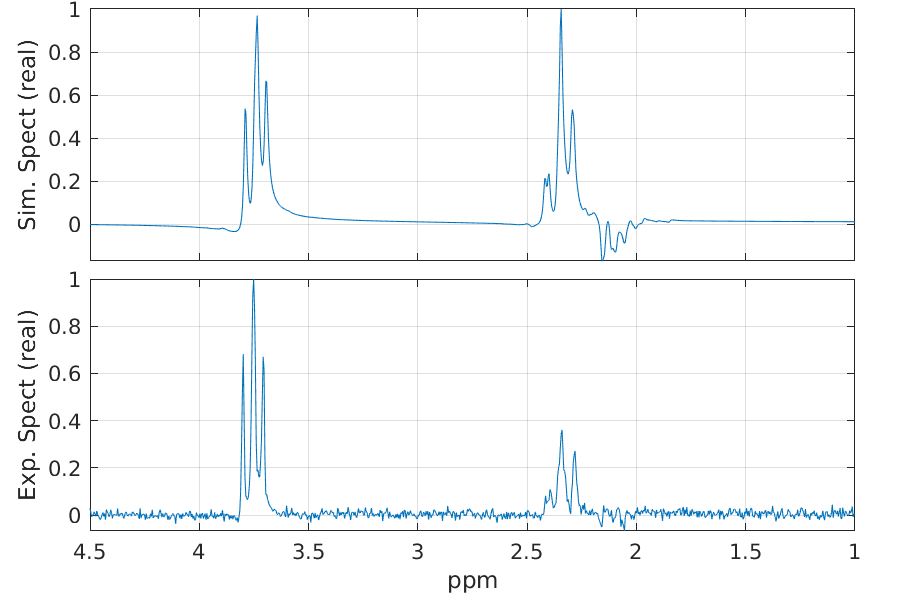}}
  \subfloat[Gln ON]{\includegraphics[width=0.195\linewidth]{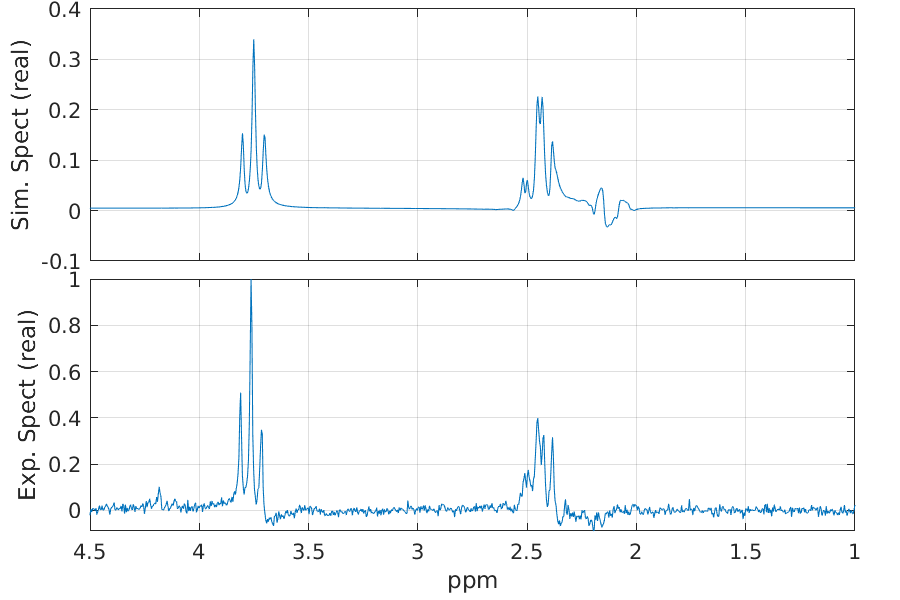}}
  \caption{Comparison of simulated (FID-A) vs experimental (Skyra) basis spectra. Experimental basis spectra derived from $\SI{100}{mM}$ metabolite solutions acquired on the same scanner.}\label{fig:compare-sim-vs-exp-basis}
\end{figure}

The performance drop demonstrates a sim-to-real gap that also affects other medical deep learning applications. Although our simulations were rather sophisticated---solving the time-dependent Schr{\"o}dinger equation for accepted Hamiltonian models of the metabolites, simulating actual pulse sequences with accurate pulse timings and pulse profiles, incorporating phase cycling and realistic slice profiles to account for spatial localisation effects---they failed to capture the full range of subtle variations present in experimentally acquired signals. Discrepancies can arise from several physical and chemical factors not fully modelled in the simulation:
\begin{itemize}
  \item \textbf{Lineshape and Phase Variations:} Minor, unmodelled variations in peak lineshapes (deviating from pure Lorentzian/Gaussian forms) and phase inconsistencies across the spectrum can distort the features DL models were trained to recognise.
  \item \textbf{Baseline Instability:} The idealised baselines in simulations do not fully account for real-world instabilities arising from factors like imperfect water suppression, eddy currents, or subtle hardware artefacts, which can obscure low-amplitude peaks.
  \item \textbf{Chemical Environment Effects:} The precise chemical environment in the phantoms (e.g., pH, temperature, ionic concentration) can cause small shifts in peak frequencies that may not be perfectly aligned with the simulation's basis set.
\end{itemize}

To further investigate the issue, we compared our simulated basis spectra with experimental spectra obtained for single metabolite solutions (approx.\ $\SI{100}{mM}$), acquired using the same pulse sequence on the same scanner. Figure~\ref{fig:compare-sim-vs-exp-basis} shows that while the spectra generally match quite well, there are some systematic differences. Most notably, the simulated ON spectrum for NAA has an additional small feature between $\SI{4}{ppm}$ and $\SI{4.5}{ppm}$ that is completely absent in the experimental spectra; none of the other metabolites exhibits a comparable feature in this region. In the simulated data, this peak is reproducible and robust to added noise, effectively providing a unique fingerprint for NAA, whereas in the experimental spectra, the corresponding region is dominated by baseline and noise. The ratio of the Cr peaks at $\SI{3.9}{ppm}$ and $\SI{3.0}{ppm}$ is also slightly higher in the simulations than in the experimental spectra, and for GABA (and to a lesser extent Glu) the triplet around $\SI{2.3}{ppm}$ appears sharper and less noise-affected in the simulated basis than in the experimental data. Together with our sim-to-real analysis on the phantom series, this suggests that the models may have learned to rely on subtle, simulation-specific features and peak patterns robust to noise that are stable in simulated data but absent, attenuated or distorted in experimental spectra.

A more detailed view is provided by the per-series sim-to-real metrics in the accompanying repository~\cite{results-sim2real}. The series with the highest quantification errors, E7, E9, and E11, also show the largest residual magnitude and phase discrepancy between simulated and experimental spectra in that analysis, consistent with the concentration-ratio and biochemical conditions (e.g., extreme GABA/NAA or Glu/NAA) being underrepresented in training and harder to match by our fixed-linewidth simulation. Decomposing the residual into distinct contributions from lineshape asymmetries, phase errors, and baseline drift would require targeted simulation experiments (e.g., fitting asymmetric lineshapes or perturbed phase to phantom spectra).

\begin{figure}[t]
  \centering
  \subfloat[Solution series (E1)]{\includegraphics[width=0.48\linewidth]{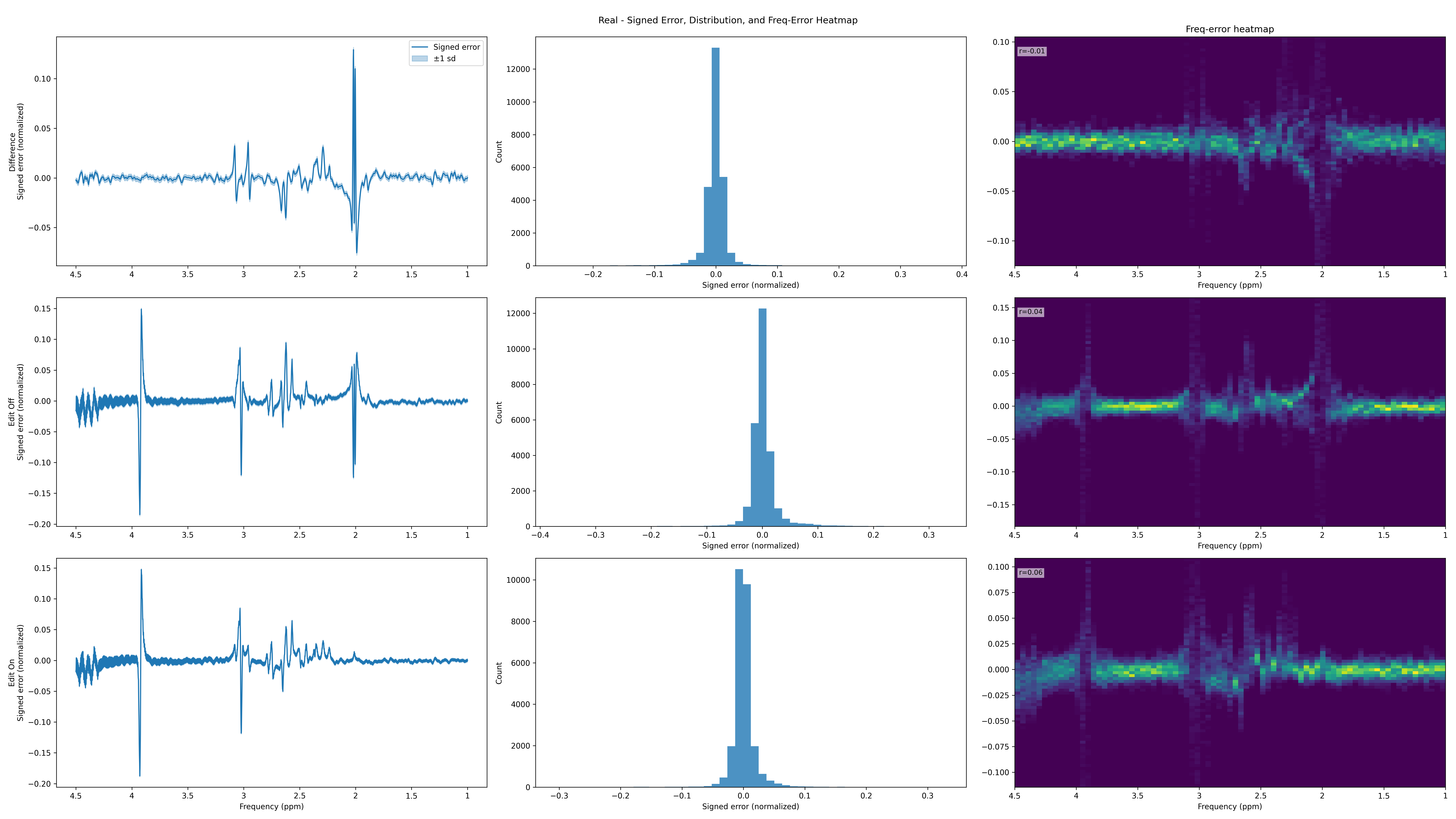}}
  \hfill
  \subfloat[Gel phantom series (E4a)]{\includegraphics[width=0.48\linewidth]{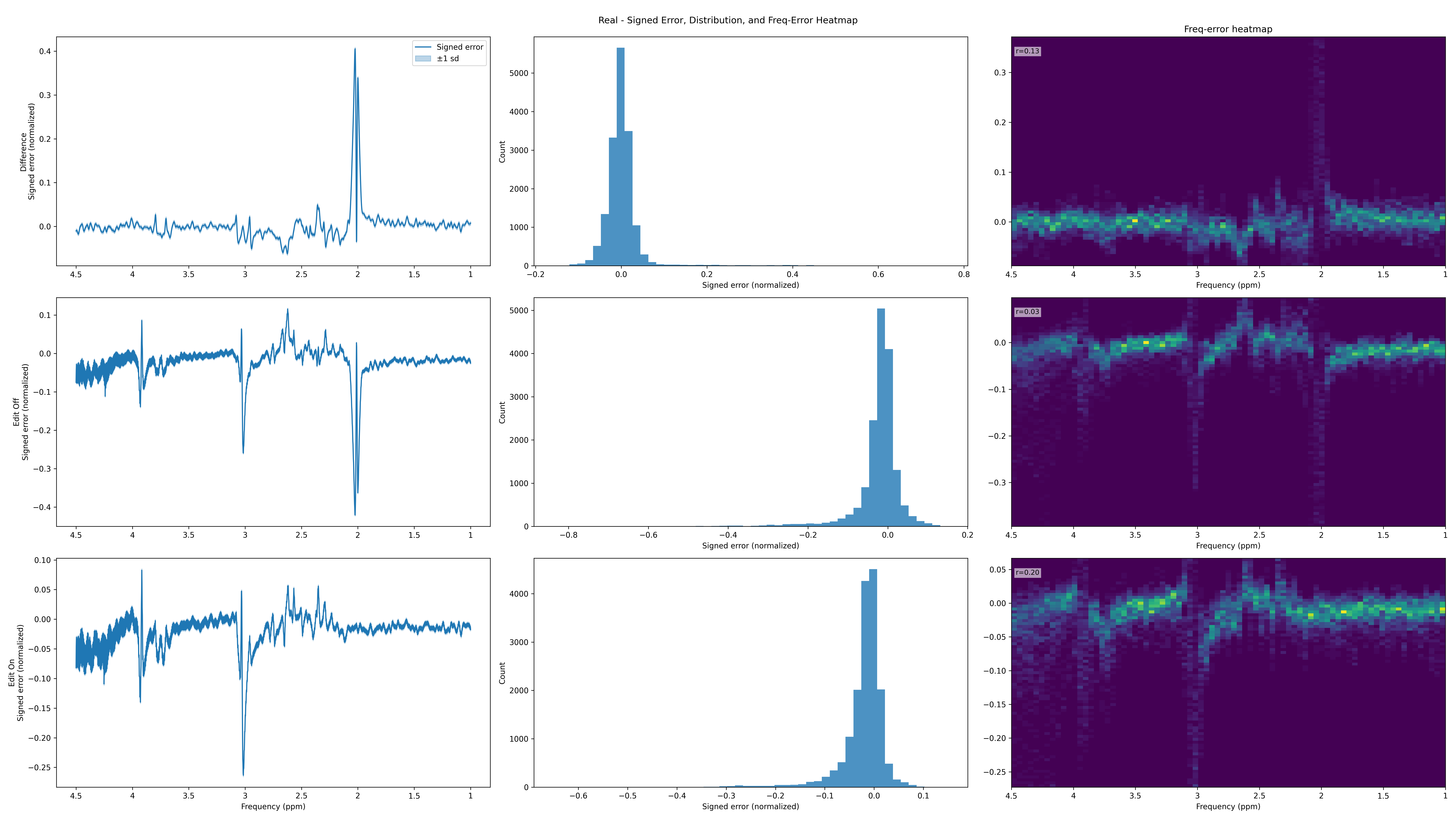}}
  \caption{Representative comparison of simulated vs experimental ON spectra for a solution series (E1) and a gel phantom series (E4a), together with their residuals. Overall spectral magnitudes and peak positions agree well, particularly for the solution series, while the gel phantoms exhibit broader lineshapes and more pronounced baseline and phase differences that are closer to \emph{in vivo} conditions.}\label{fig:sim2real-examples}
\end{figure}

\subsubsection{Interpreting Model Behaviour and Methodological Implications}

Architectural differences between our models help explain their different failure modes. The YAE, designed to learn a compressed, global representation via its autoencoder, is effective on structurally consistent simulated data. However, this global focus becomes a weakness when presented with experimental data. Baseline distortions and lineshape variations can obscure the subtle features of low-concentration metabolites, preventing them from being well-encoded in the latent space and subsequently reconstructed or quantified. This may explain its difficulty in resolving the fine details of metabolites like Gln and Cr.

The CNN architecture is biased towards extracting local features (i.e., spectral peaks) through its convolutional filters. This makes it theoretically more robust to baseline issues and more adept at identifying weak signals like Gln. However, its failure suggests that the filters were trained to recognise the overly specific, idealised peak shapes present in the simulations. They did not generalise well to the broadened or otherwise distorted peaks found in the phantom data.

The FCNN, particularly with its concatenated ReLU activation functions, performed comparably to the YAE on several metrics (e.g., best overall MAE). Its strong phantom performance should be interpreted with caution given overfitting on simulated data (Section~\ref{sec:limitations}). That said, a direct regressor can match the YAE here because the task may not need the full capacity of an autoencoder, and the FCNN avoids YAE's latent bottleneck and decoder. So under domain shift, it may rely on features that transfer better.

\subsubsection{Implications for Clinical Translation}

Acceptance of any new MRS quantification method in clinical or experimental settings, traditional or DL-based, requires phantom validation with known concentrations. The next step, where feasible, is validation on \emph{in vivo} data (where macromolecule modelling is critical) and across multiple sites and vendors to establish generalisability.

\subsubsection{Limitations}\label{sec:limitations}

This study has several limitations. Firstly, although we identified a sim-to-real gap and discussed plausible causes, we did not perform an exhaustive quantitative analysis of spectral differences between simulated and phantom data. Summary statistics from our sim-to-real comparison nevertheless indicate good magnitude agreement (correlation $0.90$, cosine similarity $0.92$), lower phase agreement (correlation $0.46$), and mean NAA and Cr linewidths (FWHM in ppm) of $0.041$ and $0.038$ and SNRs of $26.3$ and $14.3$ respectively; a more exhaustive analysis could guide improved simulations.

Secondly, validation was confined to phantoms. While phantoms are essential for ground-truth evaluation, they do not capture the full complexity of \emph{in vivo} data (e.g., macromolecules, lipids, physiological noise). Phantom validation is therefore a necessary but not sufficient step toward establishing clinical utility. In particular, our phantoms do not include a macromolecule (MM) background\@. \emph{In vivo}, the overlap of GABA with MM at $\sim\SI{3}{ppm}$ is a major challenge that edited MEGA-PRESS and difference-spectrum quantification (e.g., LCModel-DIFF) are specifically designed to address. The absence of MM in our phantom data means that the relative merits of OFF-based versus DIFF-based quantification observed may not directly transfer to human data, where MM modelling is critical; clinical readers should interpret our findings in that light. Furthermore, the effectiveness of our linewidth-augmentation strategy in \emph{in vivo} data, where MM contributes a strong overlapping signal at $\sim\SI{3}{ppm}$, may differ from that observed in phantoms; LCModel-DIFF and edited MEGA-PRESS are specifically designed to separate GABA from MM via difference spectra. This phantom validation should therefore be seen as a necessary step toward clinical deployment rather than a complete \emph{in vivo} solution.

Thirdly, our findings are specific to MEGA-PRESS and the phantom conditions and scanners used; validation was performed at a single site on a Siemens $\SI{3}{T}$ system, and generalisation to other sequences, sites, or vendors would require further validation. Furthermore, FCNN's phantom performance may be inflated by mild overfitting on simulated data.

\subsubsection{Future Work}

Reducing the sim-to-real gap requires more realistic simulations and augmentation, and possibly domain adaptation or mixed simulation--phantom training. The sim-to-real comparison (Figure~\ref{fig:sim2real-examples}) and the structure of residuals between simulated and experimental spectra indicate that simulations must capture realistic peak shapes as well as distortions and various noise sources and types. Improving model architecture alone may have limited impact if the training data do not better match experimental variability.

In parallel, more sophisticated training strategies are needed. Techniques from domain adaptation and transfer learning could help models generalise from simulation to experimental domains. Training on mixed datasets of simulated and experimental spectra is another possibility, though care must be taken to address the inherent data imbalance where a few real spectra may be ignored by the model. Semi-supervised or self-supervised learning on large-scale unlabelled \emph{in vivo} data could also improve robustness, even if the lack of ground truth remains a challenge. We provide a benchmark and argue that reliability of any new MRS quantification method should be judged against experimental phantom data with known concentrations.

\section{Conclusion}\label{sec:conclusion}

We presented a systematic evaluation of different deep learning architectures for GABA quantification from MEGA-PRESS spectra, focusing on a Y-shaped autoencoder (YAE) and a convolutional neural network (CNN), and including several adapted baseline architectures (FCNN, QNet-default, QNetBasis, QMRS and EncDec). We selected CNN and YAE configurations via Bayesian optimisation on $10{,}000$ simulated spectra (five-fold cross-validation), trained the chosen models on $100{,}000$ spectra, and evaluated them on $144$ phantom spectra with known concentrations, alongside LCModel-OFF\@.

Our principal finding is that while non-augmented deep learning models showed performance degradation on experimental phantoms, models trained with variable linewidth augmentation showed lower errors. The augmented YAE and FCNN models achieved an MAE for GABA over all phantom spectra of $0.151$ and $0.160$, respectively, surpassing the conventional LCModel-OFF baseline ($0.220$). Similar improvements were observed for Glutamate, and partial improvements for Glutamine, depending on the architecture. Table~\ref{tab:linewidth_aug_mae} reports the augmented results; Table~\ref{tab:overall_mae} gives the non-augmented comparison. The augmented DL models perform better for GABA and Glu (and FCNN for Gln), while Cr and NAA are more simply and accurately quantified by LCModel from OFF spectra.

Translating DL-based MRS quantification from simulation to clinical use remains difficult; phantom validation is needed to quantify the gap. The performance gap likely stems from subtle variations in experimental spectra (e.g., lineshape distortions, baseline instabilities, unmodelled chemical shifts) not fully captured by high-fidelity simulations. Training initially used fixed linewidth, and variable linewidth augmentation improved the accuracy on the phantom spectra. However, a notable sim-to-real gap remained, and it is unclear whether it stems from a fundamental limitation or training data fidelity. Performance gains may be achieved by improving simulation and training strategies rather than by changing model architectures. This must be combined with validation on experimental data with known ground truth to devise trusted tools in clinical and biomedical research.

\section*{CRediT Authorship Contribution Statement}

Zien Ma: Conceptualization; Methodology; Software; Validation; Formal analysis; Investigation; Data curation; Visualization; Writing --- original draft.

Oktay Karaku\c{s}: Writing --- review \& editing.

Sophie M.\ Shermer: Methodology; Resources; Data curation; Writing --- review \& editing.

Frank C.\ Langbein: Methodology; Software; Formal analysis; Data curation;  Writing --- review \& editing.

\section*{Declaration of Competing Interest}

The authors declare no competing interests.

\section*{Data Availability}

Experimental phantom compositions and processed spectra (Swansea 3\,T MEGA-PRESS phantom series E1--E14), along with summary evaluation tables and figures, are available at \url{https://qyber.black/mrs/data-megapress-spectra}. Basis spectra for simulation are at \url{https://qyber.black/mrs/data-mrsnet-basis}. Simulated datasets can be generated using scripts and configuration in the MRSNet code repository; pre-generated simulated MEGA-PRESS spectra are also available at \url{https://qyber.black/mrs/data-mrsnet-simulated-spectra-megapress}. The sim-to-real analysis (per-series metrics and comparison reports) is at \url{https://qyber.black/mrs/results-mrsnet-sim2real}~\cite{results-sim2real}. CNN, YAE, and extra model selection results are at \url{https://qyber.black/mrs/results-mrsnet-models-cnn}, \url{https://qyber.black/mrs/results-mrsnet-models-yae}, and \url{https://qyber.black/mrs/results-mrsnet-models-extra}~\cite{repo-model-cnn,repo-model-ae}. Trained models are at \url{https://qyber.black/mrs/data-mrsnet-models}. All listed repositories are tagged v2.1 for consistency and released under AGPL-3.0-or-later.

\section*{Code Availability}

Training, inference, analysis and simulation code, including model configurations, are released at \url{https://qyber.black/mrs/code-mrsnet}~\cite{MRSNET-code} (tag v2.1), with a versioned DOI at Zenodo, \url{https://doi.org/10.5281/zenodo.18520504}. DICOM reading for Siemens IMA spectroscopy files is provided by the QDicom Utilities~\cite{qdicom}. Basis spectra simulation uses FID-A~\cite{FID-A2017} (V1.2). The repository README and \texttt{requirements.txt} list further dependencies (e.g.\ Python, TensorFlow). All code is released under AGPL-3.0-or-later.

\section*{Funding and Acknowledgements}

The authors acknowledge the use of the MRI scanner facilities at Swansea University. The chemicals for the phantom studies were sponsored by Cardiff University School of Computer Science and Informatics funds.

\section*{Ethical Approval}

This study used only physical phantoms; no human participants or animals were involved. Therefore, ethical approval was not required.

\bibliographystyle{elsarticle-num}
\bibliography{ref.bib}

\clearpage
\rule{\textwidth}{.1pt}\\[1ex]
{\centering\textbf{\large Graphical Abstract}\\[1ex]
\includegraphics[width=\textwidth]{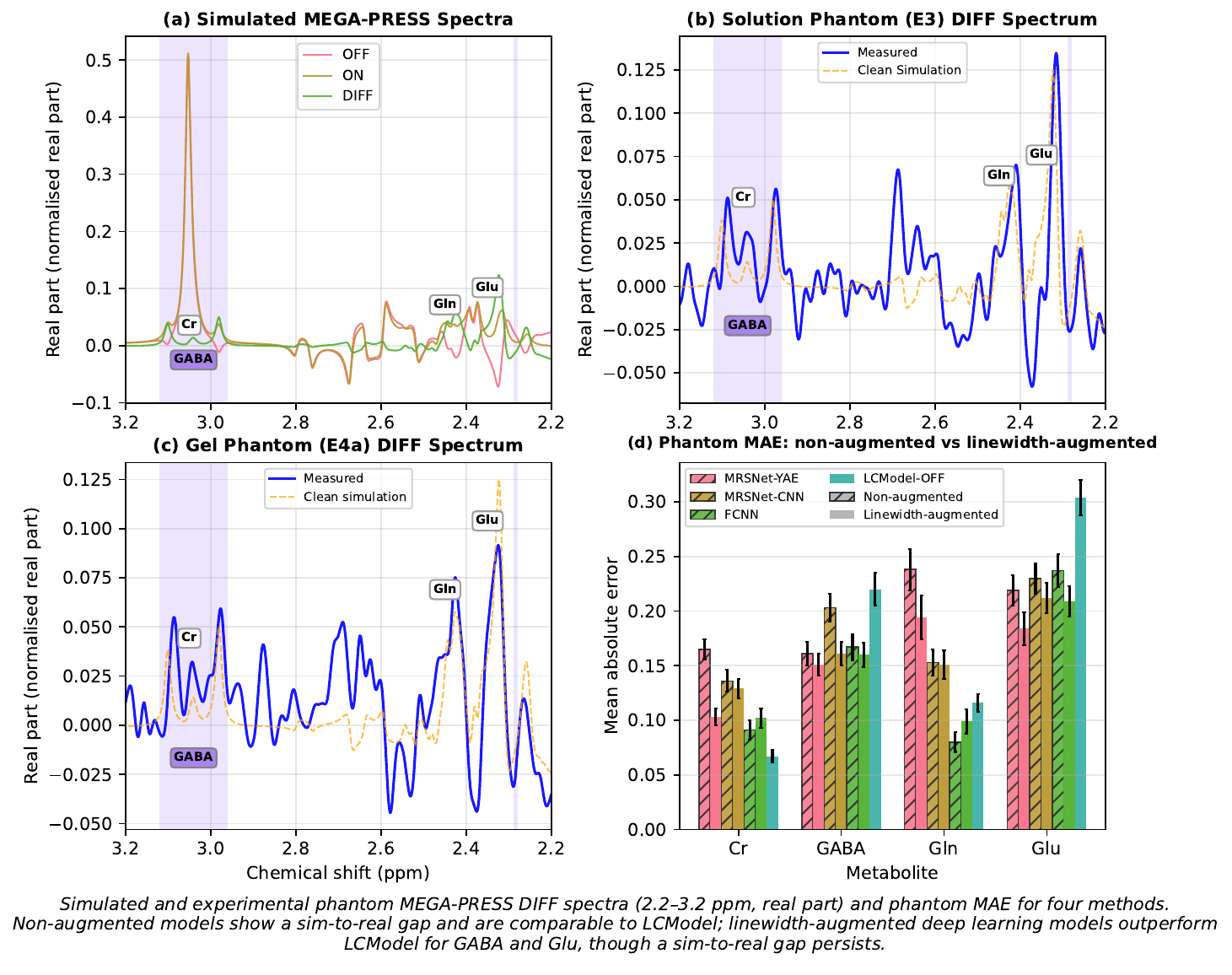}}
\rule{\textwidth}{.1pt}\\[1ex]
\textbf{Highlights}
\begin{itemize}
  \item Systematic Bayesian model selection of deep learning models on realistic simulations
  \item Phantom ground-truth validation across solution and gel series at $\SI{3}{T}$
  \item Non-augmented models show a sim-to-real gap and are comparable to LCModel
  \item Linewidth-augmented DL outperforms LCModel for GABA and Glu on phantoms; gap remains
\end{itemize}
\rule{\textwidth}{.1pt}

\end{document}